\documentstyle[12pt,aaspp4]{article}

\newcommand\et{{\it et al.~}}
\newcommand\ergs{{\rm erg ~s^{-1}}}
\newcommand\del{\partial}

\newcommand\cm{{\rm ~cm}}
\newcommand\s{{\rm s}}
\newcommand\few{{\rm few}}
\newcommand\keV{{\rm keV}}
\newcommand\msun{{\rm ~M_\odot}}
\newcommand\msyr{{\rm ~M_\odot ~yr^{-1}}}
\newcommand\pers{{\rm ~s^{-1}}}
\newcommand\ecps{{\rm ~erg~ cm^{-2}~ s^{-1}}}
\newcommand\gcm{{\rm ~g~ cm^{-3}}}

\begin{document}

\title{Accretion Disk Boundary Layers Around Neutron Stars: X-ray
Production in Low-Mass X-ray Binaries}

\author{Robert Popham and Rashid Sunyaev\altaffilmark{1}}

\altaffiltext{1}{also Space Research Institute, Moscow, Russia}

\affil{Max Planck Institut f\"ur Astrophysik}
\affil{Karl-Schwarzschild-Strasse 1, 85740 Garching, Germany, 85740;
popham@mpa-garching.mpg.de, sunyaev@mpa-garching.mpg.de}

\lefthead{POPHAM AND SUNYAEV}
\righthead{BOUNDARY LAYERS AROUND NEUTRON STARS}

\begin{abstract}

The boundary layer where the accretion disk meets the star is expected
to be the dominant source of high-energy radiation in low-mass X-ray
binaries which contain weakly magnetized accreting neutron stars.  We
present solutions for the structure of the boundary layer in such a
system.

We find that the main portion of the boundary layer gas is hot ($\ga
10^8$ K), has low density, and is radially and vertically extended.
It will emit a large luminosity in X-rays, mainly produced by
Comptonization of soft photons which pass through the hot gas.  The
gas is generally optically thick to scattering but optically thin to
absorption.  Energy is transported by viscosity from the rapidly
rotating outer part of the boundary layer to the slowly rotating inner
part, and this has the important effect of concentrating the energy
dissipation in the dense, optically thick zone close to the stellar
surface.  Advection of energy also plays an important role in the
energy balance.  We explore the dependence of the boundary layer
structure on the mass accretion rate and rotation rate of the star.
We also examine the effects of changes in the $\alpha$ viscosity
parameter and the viscosity prescription.

Radiation pressure is the dominant source of pressure in the boundary
layer.  The radiation flux in the boundary layer is a substantial
fraction of the Eddington limiting flux even for luminosities well
below ($\sim 0.01$ times) the Eddington luminosity $L_{Edd}$ for
spherically symmetric accretion.  At luminosities near $L_{Edd}$, the
boundary layer expands radially, and has a radial extent larger than
one stellar radius.  This radial expansion increases the surface area
of the boundary layer and allows it to radiate a larger total
luminosity.

Based on the temperatures and optical depths which characterize the
boundary layer, we expect that Comptonization will produce a power-law
spectrum at low source luminosities.  At high luminosities the
scattering optical depth is quite large, and bremsstrahlung and
Comptonization will produce a Planckian spectrum in the dense region
where most of the energy is released.  This spectrum will be altered
by Comptonization as the radiation propagates through the
lower-density outer boundary layer.  We discuss some implications of
our results for standard multi-component fits to X-ray spectra of
LMXBs.

\end{abstract}

\section{Introduction}

\subsection{The Boundary Layer}

The boundary layer, the region where the rapidly spinning disk
material reaches the more slowly spinning accreting star, is a crucial
element of an accretion disk.  In a thin accretion disk, the gas
rotates at approximately the Keplerian velocity, and so by the time it
approaches the surface of the accreting star, half of the
gravitational potential energy released in the accretion process has
been converted into rotational energy of the gas (in the Newtonian
case).  Unless the star is rotating rapidly, most of this energy will
be released in the boundary layer.  Furthermore, since this energy
comes from a small region close to the star, the boundary layer should
be hotter than the disk and should produce harder radiation.  Thus,
the boundary layer is expected to be the dominant source of
high-energy emission from accretion disks, which for accretion onto
neutron stars takes the form of X-rays.

The boundary layer region is quite complex, since the accreting gas
must make the transition from a disk to a star, with the accompanying
changes in the balance of momentum and energy.  For example, as the
rotation rate of the gas drops below Keplerian, rotational support
against gravity is replaced by pressure support.  The radiation which
cools the disk flows vertically, from the disk midplane to the
surface, while near the stellar surface radiation must flow radially
outward.  In general, the radial scale over which the disk properties
vary becomes comparable to or smaller than the vertical scale of the
disk.  As a result, radial transport, particularly of energy, plays an
important role in the boundary layer.

Boundary layers around neutron stars are even more complex, due to a
number of additional physical processes which become important due to
the small size of the neutron star and the resulting strong gravity
and enormous luminosity due to accretion.  In the solutions we present
here, radiation pressure plays a major role in the dynamics of the
flow, and can increase the sound speed to $\ga 0.1 c$.  Radiation
pressure is dominant even for accretion rates where the total
luminosity is well below (0.01 times) the Eddington limit, since the
local radiation flux still reaches a large fraction of the local
Eddington value.  The gas can reach very high temperatures, so
Comptonization of soft incident photons can be an important energy
loss mechanism and can produce power-law spectra (Sunyaev \& Titarchuk
1980).  Energy transfer between protons and electrons can become
inefficient at high temperatures, producing a two-temperature plasma.
Relativistic effects can also be important, since the neutron star
radius is comparable to the radius of the last stable particle orbit
in the Schwarzschild metric.  The incident radiation could in
principle remove angular momentum from the gas (Miller \& Lamb 1993,
1996), and if the neutron star radius is smaller than the last stable
orbit, the gas may spiral toward the surface (Klu\'zniak \& Wilson
1991).

\subsection{Observational Background}

Low-mass X-ray binaries (LMXBs) have been observed in X-rays for
almost 40 years, since the discovery of Sco X-1 (Giacconi \et 1962).
Since then, around 100 other LMXBs have been discovered, and many of
these have been observed extensively in X-rays and at other
wavelengths.  These observations have revealed much about the timing
and spectral behavior of LMXBs.  

The best-studied LMXBs generally fall into two classes, the atoll
sources and the Z sources, based on the paths they trace out in the
X-ray color-color diagram as they vary in brightness (Hasinger \& van
der Klis 1989).  These variations are believed to be due to changes in
the mass accretion rate $\dot M$, with the Z sources having
luminosities near the Eddington limit and the atoll sources varying
over a wider range, down to about 1\% of the Eddington limit.  If so,
then the atoll sources in particular provide an excellent means for
directly observing the effects of changing $\dot M$ and checking the
predictions of our models.

As the sources move along the paths in the color-color diagram, their
variability properties also change, as indicated by changes in the
shape of their power density spectra.  In some regimes, quasi-periodic
oscillations (QPOs) appear.  Some of these are at relatively low
frequencies of a few Hz or tens of Hz (van der Klis \et 1985;
Middleditch \& Priedhorsky 1986), but others are the recently
discovered kHz QPOs (see van der Klis 1998 for a review).  Sunyaev \&
Revnivtsev (2000) have recently shown that power density spectra of
LMXBs in the low/hard state show much more power at high frequencies
(> 100 Hz) than those of black hole candidates, probably due to the
presence of the neutron star surface and the associated energy release
in a boundary layer.  The high-frequency variability and oscillations
presumably originate from the innermost portions of the accretion
flow, very near the neutron star, and provide strong motivation for
studying these inner regions in detail.

Nonetheless, our understanding of the production of X-rays and the
formation of the X-ray spectrum in these sources is still very
incomplete.  The X-ray spectra of LMXBs are generally modeled in a
rather simple way, by combining two or more spectral components to fit
the overall spectrum.  The components consist of blackbody, ``disk''
(a sum of blackbodies corresponding to annuli of a disk with a certain
temperature and emissivity profile), bremsstrahlung, or power-law
spectra.  Thompson (elastic) scattering or Comptonization (where the
frequency changes due to scattering) can play important roles in
producing or modifying these components.

These model fits are useful in that they have provided a simple
picture of the changes in LMXB spectra as a function of luminosity.
At high luminosities, above around $10^{37} \ergs$, the data are fit
well by blackbody-type spectra.  Mitsuda \et (1984) used a
two-component model with a disk component and a single-temperature
blackbody to fit high-luminosity LMXB spectra.  White \et (1986) used
a similar model in which the disk emission was Compton-scattered to
higher energies.  At lower luminosities, the spectra are fit well by a
power-law spectrum with an exponential cutoff at high energies (White,
Stella, \& Parmar 1988).  Both a soft component and a hard power law
can be present at low luminosities, but the power law seems to
disappear at higher luminosities (Barret \& Vedrenne 1994).

The components used in these fits incorporate much of our current
knowledge of the processes by which hot gas emits X-rays.  However, in
order to better understand which components should be present and how
their temperatures and luminosities relate to each other, it is
important to have a model for the accretion flow near the neutron
star.  In particular, the boundary layer region, where the accretion
disk meets the star, is expected to produce a large portion of the
X-ray luminosity.  The size, temperature, and optical depth of this
region depend on the dynamics and energetics of the accretion flow
near the stellar surface.  In this paper, we study the boundary layer
region in detail, as a step toward the eventual goal of being able to
directly interpret LMXB spectra in terms of fundamental parameters
such as the mass accretion rate and the rotation rate of the accreting
neutron star.

\subsection{Theoretical Background}

The structure of the boundary layer region has been studied in other
types of accreting systems, most notably cataclysmic variables
(Pringle 1977; Pringle \& Savonije 1979; Tylenda 1981; Patterson \&
Raymond 1985; Kley 1991; Narayan \& Popham 1993; Popham \& Narayan
1995) and pre-main sequence stars such as T Tauri and FU Orionis stars
(Popham \et 1993, 1996).  Cataclysmic variables (CVs) are similar to
LMXBs in many respects; the main difference is simply that accretion
is onto a white dwarf instead of a neutron star.  CVs also emit X-rays
which are believed to originate in the boundary layer.  Narayan \&
Popham (1993, hereafter NP93) showed that the optical depth of the
boundary layer region is sensitive to the mass accretion rate.  At
high accretion rates, the boundary layer is optically thick and emits
approximately as a blackbody with an effective temperature of a $\few
\times 10^5$.  But at low accretion rates, the boundary layer becomes
optically thin to absorption and is unable to cool efficiently, as had
been predicted by Tylenda (1981) and King \& Shaviv (1984).  The
accreting gas is heated to $\sim 10^8$ K by the energy dissipated in
the boundary layer, and emits hard X-rays.

While we expect some similarities between boundary layers in CVs and
in LMXBs, we also expect a number of major differences.  Because a
neutron star is so much smaller than a white dwarf, a much larger
luminosity must be emitted from a much smaller area, resulting in much
higher radiation fluxes and temperatures, and therefore Comptonization
and radiation pressure play critical roles.

Studies of the accretion flow onto neutron stars have largely focused
on the case where the neutron star has a very strong magnetic field.
This field is believed to truncate the accretion disk at some inner
radius and channel the accretion onto magnetic field lines, so that
ultimately it falls onto polar caps corresponding to the poles of the
magnetic field (Pringle \& Rees 1972; Basko \& Sunyaev 1976; Ghosh,
Lamb, \& Pethick 1977).  Evidence for this magnetically-channeled
polar accretion is provided by the X-ray pulsations and magnetic
cyclotron features first observed in Her X-1 (Tr\"umper \et 1978) and
in a number of other X-ray pulsars.  However, the great majority of
LMXBs do not show any evidence for periodic pulsations or cyclotron
features.  In these systems the magnetic field may be sufficiently
small ($B \la 10^8$ G, as in some millisecond pulsars) to allow the
disk to extend all the way in to the stellar surface, resulting in a
boundary layer region where the rapidly rotating disk meets the
(presumably) more slowly rotating star.

There have only been a few studies of the inner accretion flow onto
non-magnetic neutron stars (those where the stellar magnetic field is
not strong enough to alter the flow), but for the most part they have
not computed the boundary layer structure in detail.  Sunyaev \&
Shakura (1986) computed the relative contributions of the disk and
boundary layer to the total accretion luminosity.  In standard
Newtonian disk theory each contribute half of the total, but in the
Schwarzschild metric the boundary layer contributes more than the
disk.  The relative contributions depend on the neutron star radius
$R_*$: if $R_*$ equals the radius $R_{ms}=6 G M /c^2$ of the
marginally stable particle orbit, the boundary layer luminosity should
be about twice that of the disk.  If $R_* < R_{ms}$ and the accreting
gas spirals rapidly in from $R_{ms}$ to $R_*$, the relative
contribution of the boundary layer is even larger.  The relative
luminosities of the disk and boundary layer also depend on the
rotation rate of the star, as shown by Sibgatullin \& Sunyaev (1998),
who included the effects of rotation on the shape of the star and its
gravitational field.  Shakura \& Sunyaev (1988) derived analytic
estimates for the boundary layer structure at low accretion rates and
X-ray luminosities $L_x < 10^{36} \ergs$, assuming constant
temperature and viscosity coefficient.  They did not address the case
of higher luminosities, where radiation pressure should dominate.
Klu\'zniak \& Wilson (1991) computed the structure of an accretion
belt on the stellar surface under the assumption that $R_* < R_{ms}$,
and the accreting gas impacts the surface at high velocity, and found
that high temperatures and hard spectra would be produced.  King \&
Lasota (1987) argued that the boundary layer region would reach high
temperatures even if the gas does not experience rapid infall, because
as in the CV case, the gas cannot cool efficiently enough to radiate
away the dissipated energy.  They show that for luminosities less than
$\sim 10^{35} \ergs$, where gas pressure is dominant, the boundary
layer region should heat up and expand vertically to form a ``corona''
around the neutron star.

Recently, Inogamov \& Sunyaev (1999) have studied the problem of disk
accretion onto neutron stars, using a new approach for modeling the
boundary layer region.  The accreting gas arrives at the equator of
the star spinning at the Keplerian velocity, and forms a layer on the
stellar surface which spreads from the equator toward the poles.  As
the gas moves meridionally, it loses angular momentum and dissipates
energy, which is radiated away from the surface.  This approach
essentially treats the boundary layer as part of the star rather than
part of the disk, and the angular velocity decreases with latitude on
the stellar surface rather than with radial distance from the surface.
This complements the approach used in the current paper, and later we
compare the results of the two approaches.

We model the boundary layer as part of the disk, using the slim disk
equations (Paczy\'nski \& Bisnovatyi-Kogan 1981; Muchotrzeb \&
Paczy\'nski 1982; Abramowicz \et 1988), which contain terms which
allow for large deviations from the standard thin Keplerian disk with
efficient cooling.  A similar approach has been used in most previous
studies of boundary layers in CVs and accreting pre-main sequence
stars.  This approach has the advantage that it allows one to solve
for the structure of the disk and boundary layer together, using a
single set of equations throughout.  The interface with the accreting
star is treated as a set of boundary conditions implemented at the
stellar radius.  We use Newtonian equations throughout, despite the
fact that our adopted neutron star radius of 10 km is less than the
radius of the marginally stable particle orbit (12.4 km for a neutron
star mass of 1.4 $\msun$).  Since there have been no previous
solutions of these equations for neutron star parameters, either in
Newtonian or relativistic form, we feel that a Newtonian solution is
an important first step.  The effects of relativity will be added in
the future.

Figure 1 shows some of the important features of our results.  At the
transition from the disk to the boundary layer, the angular velocity
$\Omega$ reaches a maximum and the flow passes through a narrow neck,
where the disk height is only $\sim 40 - 70$ meters in the low-$\dot
M$ solutions.  The radial extent of the boundary layer region is $\sim
0.1-0.2$ of the stellar radius at low $\dot M$, and the height is
comparable to the radial extent.  The angular velocity drops slowly
over most of the boundary layer, and then rapidly at the inner edge.
At high $\dot M$ near the Eddington limit, the situation is quite
different: the radial extent and height boundary layer are equal to
the stellar radius, and the neck between the disk and boundary layer
is much wider.  Here the drop in angular velocity occurs over the
whole width of the layer.

In \S 2, we describe the slim disk equations and radiative transfer
scheme which we have used to model boundary layers in LMXBs.  We
present expressions for the viscous transport of energy in the disk
and boundary layer.  We present the results of our calculations in \S
3, and show how the transport of energy by viscosity, radiation, and
advection play essential roles in determining the boundary layer
structure.  We present solutions for a variety of mass accretion
rates, stellar rotation rates, and viscosities, and show how the size,
temperature, and other properties vary.  In \S 4 we discuss the
energetics of the boundary layer, the behavior near the Eddington
limit, and the implications of our results for the spectra of LMXBs.

\section{Boundary Layer Model}

\subsection{Dynamics}

We use the slim disk equations to describe accretion in the disk and
boundary layer.  Like the standard thin disk equations, these describe
the disk structure as a function of radius.  In the vertical
direction, we assume simple approximate relations.  The slim disk
equations are a generalization of the standard thin disk equations
which include terms that become important when the disk deviates from
a thin, Keplerian configuration.  These additional terms often involve
radial derivatives.  The equations are solved assuming a steady state,
and the solutions extend from the radius of the stellar surface
(assumed to be 10 km) out to 100 times the stellar radius.  Boundary
conditions for the flow are set at both the inner and outer radii, and
the equations are solved using a relaxation method.  

The slim disk equations have been presented in a number of previous
papers, and therefore we only give a brief description of them here;
for more details consult, e.g. Popham \& Narayan (1995, hereafter
PN95).  The mass accretion rate through the disk, which is constant
with radius under the steady-state assumption, is given by
\begin{equation}
\dot M = - 4 \pi R H \rho v_R,
\end{equation}
where $\dot M$ is the mass accretion rate, $R$ is the radius, $H$ is
the disk vertical scale height, $\rho$ is the mass density, and $v_R$
is the radial velocity.

The material in the disk rotates with angular velocity $\Omega$.
Viscosity transfers angular momentum down the gradient of $\Omega$,
i.e. from an annulus with higher $\Omega$ to an adjacent annulus with
lower $\Omega$.
In a steady-state disk, the flow of angular momentum is given by
\begin{equation}
\dot M {d \over dR}(\Omega R^2) = {d \over dR}(4 \pi R^2 H w_{r \phi}),
\end{equation}
where $w_{r \phi} = \rho \nu R d \Omega / dR$ is the viscous stress,
and $\nu$ is the viscosity coefficient.  This can be integrated to
obtain 
\begin{equation}
\dot M {\nu \over v_R} {d \Omega \over dR} R^2 = \dot M \Omega
R^2 - \dot J. 
\end{equation}
The integration constant $\dot J$ is the angular momentum accretion
rate.  We expect that there will be a maximum in $\Omega$ at a point
close to the surface of the star where the Keplerian disk ends and the
boundary layer begins, and $\dot J$ must be equal to $\dot M \Omega
R^2$ at this point since $d \Omega / dR = 0$.  Therefore we write
\begin{equation}
\dot J \equiv j \dot M \Omega_K (R_*) R_*^2
\end{equation}
where $R_*$ is the radius of the neutron star, and $\Omega_K (R_*)
\equiv (GM/R_*^3)^{1/2}$ is the Keplerian angular velocity at $R$.
The parameter $j$ is the ratio of $\dot J$ to the usual value $\dot M
\Omega_K (R_*) R_*^2$ assumed in the thin disk equations, where the
radial extent of the boundary layer is assumed to be very small.
Since we wish to obtain the thickness of the boundary layer from our
solutions, we allow $j \neq 1$.

The radial momentum equation is 
\begin{equation}
v_R {d v_R \over dR} + {1 \over \rho} {dP \over dR} - {1 \over
\rho}{d \over dR}\left( \rho \nu {d v_R \over dR} \right) = (\Omega^2
- \Omega_K^2) R, 
\end{equation} 
where $P$ is the total pressure.  The terms on the left-hand side of
this equation represent the radial acceleration of the accreting gas,
radial pressure gradient, and the radial viscous acceleration, all of
which can be important in the boundary layer.

The standard energy equation, which assumes that the ions and
electrons have the same temperature, is
\begin{equation}
\rho H v_R T_c {dS \over dR} + {1 \over R}{d \over dR}(R H F_R) =
\rho H \nu \left(R {d \Omega \over dR} \right)^2 - F_V,
\end{equation} 
where $T_c$ is the temperature at the disk midplane, $S$ is the
entropy, and $F_R$ and $F_V$ are the radial and vertical radiative
fluxes, respectively.  The terms on the right-hand side of this
equation are the viscous dissipation in the disk and the radiation
from the disk surface; these are assumed to be equal at all radii in
the standard thin disk equations.  The terms on the left-hand side
represent radial transport of energy and are not included in the thin
disk equations.  The first is the entropy advected inward with the
accreting gas; note that in an advection-dominated disk this term
approximately balances the viscous dissipation.  The second is
transport by radial radiation flux, which is generally quite important
in disk boundary layers.

The energy equation given above applies when the energy transfer from
ions to electrons via Coulomb collisions proceeds on a shorter
timescale than the other heating and cooling processes.  This may not
be true in the boundary layer region, where the gas may become hot and
rarefied.  Therefore we use separate energy equations for the ions and
the electrons.  The ions are heated by the energy dissipation in the
disk and cooled by Coulomb collisions with the electrons
\begin{equation}
\rho H v_R T_i {dS_i \over dR} = \rho H \nu \left(R {d \Omega \over
dR} \right)^2 - Q_{pe} H,
\end{equation} 
where $Q_{pe} = (3/2) n_p k (T_i - T_e) / \tau_{pe}$ is the Coulomb
cooling rate, and $\tau_{pe} = 2.11 \times 10^{-23} T_e^{3/2} / \rho$
s (Spitzer 1962) is the Coulomb energy transfer timescale, with $\rho$
in cgs units and $T_e$ in degrees Kelvin.  The ion temperature $T_i$
may be substantially higher than the electron temperature $T_e$.  The
electrons are heated by the Coulomb collisions, and the electron
energy equation also includes the radiation
\begin{equation}
\rho H v_R T_e {dS_e \over dR} + {1 \over R}{d \over dR}(R H F_R) =
Q_{pe} H - F_V.
\end{equation} 

The total pressure is the sum of the gas and radiation pressures
\begin{equation} 
P = {\cal R} \rho (T_i + T_e) + {4 \pi \over 3 c} u_c,
\end{equation} 
where we have assumed the gas is ionized hydrogen, ${\cal R}$ is the
gas constant, and $u$ is the mean radiative intensity as defined
below.  The entropy advection terms for the electrons and ions are
given by
\begin{eqnarray} 
T_e dS_e = {\cal R} T_e \left({3 \over 2} d \ln T_e -
d \ln \rho \right) + {4 \pi \over c \rho}\left( du_c - {4 \over 3} u_c
d \ln \rho \right),\\
T_i dS_i = {\cal R} T_i \left({3 \over 2} d \ln T_i -
d \ln \rho \right),
\end{eqnarray} 
respectively, where we have included the radiation terms in the
equation for the electrons.

We estimate the vertical pressure scale height of the disk in the
usual way.  The vertical pressure gradient must balance the vertical
component of the gravity of the star
\begin{equation}
-{1 \over \rho}{dP \over dz} = {G M \over R^2}{z \over R} =
z \Omega_K^2(R)
\end{equation}
Taking $P = \rho c_s^2$, i.e., using $c_s = (P/\rho)^{1/2}$ to define
the approximate sound speed, and assuming that the disk is isothermal
so that $c_s (z)$ is constant, we find $\rho (z) = \rho(0)
\exp(-\Omega_K^2 z^2 / 2 c_s^2)$.  Therefore we define the vertical
pressure scale height as $H = \sqrt{2} c_s / \Omega_K$.
Alternatively, if we assume that the pressure is dominated by
radiation, we have $-(1/\rho) dP/dz = \kappa F_V / c = \Omega_K^2(R)
z$, where $\kappa$ is the opacity and $F_V$ the vertical radiation
flux.  As discussed later, we assume that the vertical flux increases
linearly with $z$, $F_V (z) = F'z$, with $F'$ constant.  This gives a
quadratic variation for the pressure: $P(z) = P(0) - \rho \kappa F'
z^2 / 2 c$, with $P$ reaching zero at $z = (2 P(0) c / \rho \kappa
F')^{1/2} = \sqrt{2} c_s / \Omega_K$, which is what we have defined as
$H$.

\subsection{Viscosity and Turbulence in the Boundary Layer}

The origin and nature of the viscosity in the boundary layer may be
very different from that in the Keplerian disk.  The viscosity may
arise from turbulence, but the origin of this turbulence is not known.
There are several well-known theoretical and experimental results on
the generation of turbulence which may be relevant to the conditions
within the boundary layer.  Two of the best known mechanisms should
{\it not} operate in the boundary layer:

1) The boundary layer is stable against linear hydrodynamic
instabilities according to the Rayleigh criterion, since the specific
angular momentum of the matter increases strongly with radius within
the boundary layer, $d (\Omega R^2) / dR > 0$ (Rayleigh 1916).

2) The leading candidate for producing the viscosity in the disk is the
magnetorotational instability originally discovered by Velikhov (1959)
and Chandrasekhar (1960) and applied to accretion disks by Balbus and
Hawley (1997 and references therein).  However, this instability
arises when $d \Omega^2 /dR < 0$, which is true in the Keplerian disk,
but not true in the boundary layer.  

Nonetheless, there are other ways in which turbulence within the
boundary layer might be generated.  Two of these, which are probably
closely related, are:

1) Experimental studies of flow between two rotating cylinders have
shown that turbulence can arise for high Reynolds numbers in the case
where the outer cylinder rotates much faster than the inner one (see,
e.g., Schlichting 1957; Joseph 1976 and references therein).  This
case is not relevant to the main Keplerian portion of the disk, but
might be relevant to the boundary layer.\footnote{A possible exception
to this would be the case where the neutron star is rotating rapidly;
the fastest measured rotation frequencies of millisecond pulsars and
X-ray bursters are around 600 Hz, corresponding to angular velocities
of $\sim 1/3$ of $\Omega_K (R_*)$.}  Therefore we need to estimate the
Reynolds number $Re$ in the boundary layer.  For the boundary layer
around an accreting neutron star, the radiative viscosity $\nu_{rad} =
(4/15)(U_{rad}/\kappa_s \rho^2 c)$ (Weinberg 1972), where $U_{rad}$ is
the radiative energy density, should be well in excess of the
molecular viscosity, as discussed by Inogamov \& Sunyaev (1999), and
in the absence of turbulence the radiative viscosity would be
dominant.  We can calculate {\it a posteriori} the Reynolds number $Re
= \Omega R (R-R_*) / \nu_{rad}$ in our boundary layer solutions based
on the radiative viscosity.  We find rather low values $Re \sim 10^2$
over much of the boundary layer, but much larger values $Re \sim 10^6
- 10^{11}$ near the inner and outer edges (Fig. 2).

2) If the radial extent of the boundary layer is small compared to the
radius, $\Delta R \ll R$, then we can neglect the curvature in the
first approximation, and consider the flow to be similar to flow near
a wall.  In this situation, we know from experiments that turbulence
may arise at high Reynolds numbers.  For $\dot M = 10^{-9} \msyr$, the
radial extent of the boundary layer is $\Delta R \ll R$, and this
approximation may be valid; however, for $\dot M = 10^{-8} \msyr$ we
have $\Delta R \sim R$, and we cannot neglect the curvature of the
flow.  This case is probably very similar to the previous one, but
there is much more experimental evidence available.

Since we are treating the boundary layer as the inner part of the
accretion disk and using disk equations to describe it, we have chosen
the simple approach of applying the same viscosity law in the disk and
the boundary layer.  We use an $\alpha$ viscosity (Shakura \& Sunyaev
1973); however, in defining the viscosity coefficient, we also use the
radial pressure scale height $H_r = P / |dP/dR|$ (Papaloizou \&
Stanley 1986).  The viscosity coefficient is then defined as $ \nu =
\alpha c_s l_{turb}$, where $l_{turb} = (H^{-2} + H_r^{-2})^{-1/2};$
this essentially takes the turbulent length scale $l_{turb}$ to be the
smaller of the two scale heights.  We do not include any reduction in
the viscosity due to causality (Narayan 1992; Narayan, Loeb \& Kumar
1994); however, we find that the radial velocity stays well below the
sound speed in our solutions, so that causal corrections are
unimportant.

Another limit on the viscosity is the criterion that the length scale
for the turbulence should be short enough that the change in azimuthal
velocity $\Delta v_\phi$ over that length should be less than the
sound speed (Shakura \& Sunyaev 1988).  One can thus define a third
length scale $H_s \equiv c_s / |dv_\phi/dR|$ which should limit the
turbulence.  Since we want the viscosity to be limited by the smallest
of the three scales $H$, $H_r$, and $H_s$, we can adopt a similar
prescription to the one above, where $l_{turb} = (H^{-2} + H_r^{-2} +
H_s^{-2})^{-1/2}$.  We will refer to this prescription as ``subsonic''
viscosity.

At the mass accretion rates and luminosities we consider, the gas is
radiation pressure dominated.  Since our sound speed is based on the
total pressure including both gas and radiation pressure, it is
important that the gas and the radiation are coupled.  As long as the
optical depth across the turbulent length scale is $\ga 1$, the gas
and radiation should be reasonably well coupled.  Since the turbulent
length scale is $\sim H$, and since scattering dominates the opacity,
this amounts to the condition that the disk be optically thick to
scattering.  As we will see later, all of our solutions satisfy this
criterion, although at the lowest mass accretion rates which we
consider, the optical depth of the boundary layer from the midplane to
the surface is only a few.  If the gas density were to drop too low,
so that the disk became optically thin, the radiation would stream
freely through it, and we would need to treat the turbulence in a more
detailed way.

\subsection{Radiative Transfer}

We also need to compute the radiative fluxes in the vertical and
radial directions, $F_V$ and $F_R$, which appear in the energy
equation.  The radiative transfer in the boundary layer region can be
rather complicated due to the rapid variations in the gas temperature
and density.  In the case of LMXBs the situation is further
complicated by Compton scattering when the gas reaches high
temperatures.

We use a simple scheme for describing the radiative transfer in the
boundary layer region, in the spirit of the dynamical equations
described above.  In particular, we make simplifying assumptions about
the vertical dependence of the variables, and ignore the frequency
dependence of the radiation and opacity.  In most respects, our
equations are the same as those used by PN95 and described in detail
in Appendix A of that paper.  However, we have modified the transfer
equations to include an approximate treatment of Compton scattering.

We use a ``four-stream'' treatment.  The eight intensities
corresponding to the directions of the corners of a cube reduce to
four because we assume axisymmetry: $I^{++}, I^{+-}, I^{-+}, I^{--}$,
where the superscripts refer to the radial and vertical directions,
respectively.  We define four moments of these intensities:
\begin{mathletters}
\begin{eqnarray}
   u & = & {1 \over 4}(I^{++} + I^{+-} + I^{-+} + I^{--}) \\
   v_x & = & {1 \over 4 \sqrt{3}}(I^{++} + I^{+-} - I^{-+} - I^{--}) \\
   v_z & = & {1 \over 4 \sqrt{3}}(I^{++} - I^{+-} + I^{-+} - I^{--}) \\
   w & = & {1 \over 4}(I^{++} - I^{+-} - I^{-+} + I^{--})
\end{eqnarray}
\end{mathletters}
which correspond to the mean intensity, the radial and vertical
fluxes, and a cross term, respectively.

In place of the usual transfer equation for each of the intensities,
we use a modified version with a Compton energy amplification term
included, e.g.,
\begin{equation}
{1 \over \sqrt{3}}{d I^{++} \over dx} + {1 \over \sqrt{3}}{d I^{++}
\over dz} = -\rho \kappa I^{++} + \rho \kappa_s u (1 + \delta) + \rho
\kappa_a B, 
\end{equation} 
where $\kappa, \kappa_s, \kappa_a$ are the total, scattering, and
absorptive opacities, $\delta$ is the energy amplification factor due
to Compton scattering, and $B = \sigma T_e^4/ \pi$ is the mean
intensity of blackbody radiation at the electron temperature $T_e$.
The amplification factor is written as $\delta = (T_e -
T_{phot})/T_{com}$, where $T_{com} = m_e c^2 / 4 k \simeq 1.5 \times
10^9$ K.  The ``photon temperature'' $T_{phot}= T_e \bar \epsilon /
\bar \epsilon_{BB},$ where $\bar \epsilon$ is the mean
photon energy and $\bar \epsilon_{BB} \simeq 2.7 kT_e$ is the mean
photon energy of a blackbody distribution.  We define $\bar \epsilon =
u / N$, where $N$ is the mean photon number intensity (photons
$\cm^{-2}$ sr$^{-1} \pers$).  We calculate $N$ using a set of
equations similar to those above, but with number intensities instead
of energy intensities, the blackbody mean number intensity $N_B = B /
\bar \epsilon_{BB}$ taking the place of $B$, and no Compton
enhancement factor $\delta$, since Compton scattering conserves photon
number.

When we combine the transfer equations for the four directions, we
obtain four differential equations
\begin{mathletters}
\begin{eqnarray}
   {\del v_x \over \del x} + {\del v_z \over \del z} 
   & = & \rho \kappa_a (B - u) + \rho \kappa_s u \delta \\
   {\del u \over \del x} + {\del w \over \del z}
   & = & - 3 \rho \kappa v_x \\
   {\del w \over \del x} + {\del u \over \del z}
   & = & - 3 \rho \kappa v_z \\
   {\del v_z \over \del x} + {\del v_x \over \del z}
   & = & - \rho \kappa w.
\end{eqnarray}
\end{mathletters}

These are identical to the equations without a Compton factor
(eqs. A2.9--A2.12 of PN95), except for the second term on the
right-hand side of the first equation, which gives the Compton
scattering contribution to the total flux divergence.

We apply boundary conditions and assume vertical dependences as in
PN95; namely, $v_z$ and $w$ are zero at the disk midplane ($z=0$) and
increase linearly with $|z|$ at rates $v'_z$ and $w'$, and $u$ and
$v_x$ have their maximum values $u_c, v_{xc}$ at the midplane, and
decrease quadratically with $|z|$.  At the surface ($z=H$) the
incoming intensities $I^{+-}, I^{--}$ are assumed to be zero, so $u =
\sqrt{3} v_z$ and $w = \sqrt{3} v_x$.  We also convert the
$x$-derivatives to $R$-derivatives, and find
\begin{mathletters}
\begin{eqnarray}
	{1 \over R H}{\del (RH v_{xc}) \over \del R} + v'_z
	& = & \rho \kappa_a (B - u) + \rho \kappa_s u \delta \\
	{\del u \over \del x} + w'
	& = & - 3 \rho \kappa v_x \\
	{\del w' \over \del R} + {3 \tau + 2 \sqrt{3} \over H} v'_z
	& = & {2 u_c \over H^2} \\
	{\del v'_z \over \del R} + {\tau + 2 / \sqrt{3} \over H} w'
	& = & {2 v_{xc} \over H^2},
\end{eqnarray}
\end{mathletters}
where $\tau = \kappa \rho H$ is the vertical optical depth.
	
\subsection{Viscous Energy Transport}

A very important term in the energy balance of accretion disks is the
energy transport due to viscosity.  This energy transport is an
unavoidable consequence of the viscous angular momentum transport that
makes accretion possible.

\subsubsection{In the Boundary Layer}

The local rate of viscous dissipation of kinetic energy into thermal
energy per unit surface area of the disk is given by
\begin{equation}
Q^+ = 
w_{r \phi} H R {d \Omega \over dR} = \rho \nu H R^2 
\left( {d \Omega \over dR} \right)^2,
\end{equation}
(Shakura \& Sunyaev 1973).  This expression is a specific case of the
general formula for viscous energy dissipation (Landau \& Lifshitz
1959).  Using the continuity and angular momentum equations (1) and
(3), we can write this in the form
\begin{equation}
Q^+ = -{\dot M \over 4 \pi R}(\Omega R^2 - j \Omega_K(R_*)
R_*^2) {d \Omega \over dR}.
\end{equation}

The ultimate source of the dissipated energy is the gravitational
potential energy released as the gas falls in toward the star
\begin{equation}
Q_{grav} = {\dot M \over 4 \pi R} {d \over dR} \left( - G M \over R
\right) = {\dot M \over 4 \pi R}{G M \over R^2} = {\dot M \over 4 \pi
R} \Omega_K^2 R.
\end{equation}
However, if the radial extent of the boundary layer is very small,
$\Delta R \ll R_*$, the gravitational energy release within the
boundary layer is only a small fraction $\Delta R / R$ of the total
accretion luminosity.  Practically all of the energy dissipated in the
boundary layer then comes from the kinetic energy of the gas, which is
lost as the rotational velocity decreases from Keplerian to zero (for
a non-rotating star).  The kinetic energy is lost at a rate
\begin{equation}
Q_{kin} = {1 \over 4 \pi R} {d \over dR} \left( {1 \over 2} \dot M
\Omega^2 R^2 \right ) \simeq {\dot M \over 4 \pi R} \left(\Omega R^2
{d \Omega \over dR} \right),
\end{equation}
per unit disk surface area, where the last expression assumes that the
boundary layer is narrow so $\Omega$ varies much more rapidly than
$R$.

If the boundary layer is wide, with $\Delta R \sim R_*$, then
gravitational energy release within the boundary layer becomes
important.  Also, our expression for the rate of kinetic energy loss
must include a term proportional to $\Omega^2 R$ to account for the
change in radius. The resulting total gravitational and kinetic energy
loss rate per unit disk surface area is
\begin{equation}
Q_{grav} + Q_{kin} = {\dot M \over 4 \pi R} \left( \Omega_K^2 R +
\Omega R^2 {d \Omega \over dR} + \Omega^2 R \right).
\end{equation}
Both this and the expression in Eq. 20 differ from the local
dissipation rate given in Eq. 18.

The reason for this is the viscous transport of energy, which
redistributes energy within the boundary layer.  Viscosity causes both
angular momentum and energy to be carried down the local gradient in
$\Omega$, i.e. from regions of higher $\Omega$ to regions of lower
$\Omega$.  In the Keplerian disk, this means that angular momentum and
energy are carried outward.  At the dividing line between the disk and
the boundary layer, $\Omega$ reaches its maximum value and $d \Omega /
dR = 0$.  In the boundary layer, $\Omega$ decreases inward, so
viscosity carries both angular momentum and energy inward toward the
stellar surface.

The energy transported by viscosity is given by $\Omega N$, where $N$
is the torque
\begin{equation}
\Omega N = -4 \pi R^2 H w_{r \phi} \Omega = -4 \pi R H \rho \nu R^2 {d
\Omega \over dR} \Omega = \dot M \Omega {\nu R^2 \over v_R} {d \Omega
\over dR}.
\end{equation}
Note that this expression is proportional to $d \Omega / dR$; as
mentioned above, energy is carried down the angular velocity gradient
together with the angular momentum.  Also, energy is not carried
across a maximum in $\Omega$, such as the boundary between the disk
and boundary layer.  Therefore viscous transport {\it does not
transfer energy between the disk and boundary layer}, but
redistributes it {\it within} the disk and the boundary layer.  We can
rewrite the viscous energy transport rate using Eq. 3 as 
\begin{equation}
\Omega N = \dot M (\Omega^2 R^2 - j \Omega \Omega_K(R_*) R_*^2).  
\end{equation}
The energy per unit disk surface area deposited at a given radius is
given by the divergence of the viscous energy transport rate
\begin{equation}
Q_{trans} = -{1 \over 4 \pi R} {d \over dR} (\Omega N) = -{\dot M
\over 4 \pi R}{d \over dR} (\Omega^2 R^2 - j \Omega \Omega_K(R_*)
R_*^2).
\end{equation}
This energy is the main reason for the difference between the local
viscous dissipation rate (Eq. 18) and the local gravitational and
kinetic energy release (Eq. 20 or 21).  Other terms -- the energy
associated with the change in pressure, the bulk viscous dissipation,
and the kinetic energy of radial motion (i.e., terms proportional to
those on the left-hand side of Eq. 3) -- can also play a role; we have
omitted these here for simplicity.

In the boundary layer, $\Omega$ decreases inward, and so the viscous
transport removes energy from the outer boundary layer and carries it
inward toward the star.  The gradient $d \Omega / dR$ is much larger,
by a factor $\ga R_* / \Delta R$, in the boundary layer than in the
disk.  In addition, if $\Omega$ does not decrease uniformly through
the whole boundary layer, but instead drops more rapidly in some
sections than in others, viscous transport will be even stronger in
those sections.  Overall, we expect that viscous transport should be
very important in determining the energy balance and overall structure
of the boundary layer.  This expectation is verified by the solutions
presented in \S 3.

\subsubsection{In the Disk}

The expressions given above apply to the Keplerian disk as well.  In a
Keplerian disk, half of the gravitational energy released goes into
the rotational kinetic energy of gas.  This kinetic energy is released
later in the boundary layer.

As in the boundary layer, the local energy dissipation in the disk
does not match the local rate of gravitational and kinetic energy
release.  For example, taking $\Omega = \Omega_K$ in Eq. 18 gives $
Q^+ = (3/8 \pi) \dot M \Omega_K^2 [1 - j (R_* / R)^{1/2}]$, while the
total gravitational and kinetic energy loss rate per unit disk surface
area is is $(1/8 \pi) \dot M \Omega_K^2 $.  This has the well-known
result that at radii $R \gg R_*$, the dissipation rate is 3 times
larger than expected (see Shakura \& Sunyaev 1973).

If we take $\Omega = \Omega_K$ in Eq. 24 above, we find that the
energy per unit disk surface area deposited by viscous transport is
$Q_{trans} = (1/4 \pi)[1 - (3/2) j (R_*/ R)^{1/2}] \dot M \Omega_K^2$.
This is exactly the difference between the actual dissipation rate and
the expected dissipation rate based on the change in gravitational
potential and kinetic energy.  The energy deposited by viscous
transport is positive for $R > (9/4) j^2 R_*$ and negative for smaller
radii; energy is removed from the inner region of the disk, and the
dissipation rate there is much lower than the rate of gravitational
potential and kinetic energy loss.  Note, however, that none of this
energy comes from the boundary layer, since viscous transport cannot
carry energy across the maximum in $\Omega$.

\subsubsection{Importance for the Spectrum}

In \S 3 we calculate the rate of viscous transport in our boundary
layer solutions, and find that it plays an important role in
determining the boundary layer structure.  Most importantly, the
viscous transport causes most of the kinetic energy of the gas to be
dissipated in the densest part of the boundary layer.  There the gas
is rotating slowly and the radiation flux is well below Eddington, so
centrifugal and radiation forces do not support the gas against
gravity.  The density increases rapidly as the gas piles up on the
stellar surface.  This is crucial because it will make the radiation
spectrum quite different from that which would be produced if all the
energy were dissipated in the low-density regions farther outside.

\section{Results}

We begin by showing a typical solution with an intermediate mass
accretion rate $\dot M = 10^{-9} \msyr$, the viscosity parameter
$\alpha=0.1$, and the stellar rotation rate $\Omega_*=0$.  These three
quantities characterize our solutions.  We use a neutron star mass of
$1.4 \msun$ and a neutron star radius of $10^6 \cm$ for all of our
calculations.  Thus the accretion luminosity is for this solution is
$L_{acc} = G M \dot M / R_* = 1.17 \times 10^{37} \ergs$.  Some
equations of state for neutron stars predict a slightly larger radius,
and if the neutron star is rotating, the star will be flattened and
the equatorial radius will increase.

In discussing this solution, we begin with the disk and follow the
flow of the accreting gas inward through the boundary layer and onto
the star.  In general, the disk portion of our solutions closely
resembles the standard thin disk solutions (Shakura \& Sunyaev 1973).
The additional ``slim disk'' terms in the equations are small.

The boundary layer region can be seen clearly in Fig. 3.  As the gas
flows in, $\Omega$ begins to deviate from Keplerian at $R \simeq 1.19
\times 10^6 \cm$.  In fact, one can see in the inset that the rotation
first becomes super-Keplerian\footnote{This small super-Keplerian
region seems to be present in most of our solutions, and is due to the
drop in pressure as the gas moves from the disk into the boundary
layer.  The exceptions are solutions with high values of $\dot M$,
where the pressure increases from the disk to the boundary layer, and
those in which the star is spinning at nearly breakup speed, so that
no boundary layer is present.  It has been argued that such a region
is a necessary condition for the onset of an advection-dominated
region in a disk (Abramowicz, Igumenshchev, \& Lasota 1998).}  at $R
\simeq 1.19 \times 10^6 \cm$, then drops back below Keplerian at $R
\simeq 1.17 \times 10^6 \cm$.  The pressure drops as the gas moves
from the disk into the boundary layer (Fig. 4a), and the resulting
inward pressure gradient produces the small super-Keplerian zone.

At this radius, the gas infall accelerates rapidly; $v_R$ increases by
more than two orders of magnitude (Fig. 3), so the surface density of
the disk drops accordingly.  Note that the sound speed $c_s$ is shown
as a dashed line in Fig. 3; in the hot boundary layer region, the
radial Mach number $v_R / c_s$ is larger than in the disk, which
indicates that advection of energy becomes important there.
Nonetheless, $v_R \ll c_s$ in all of our solutions, so a causal
viscosity prescription would make little difference.

The sudden drop in surface density produces a fairly abrupt transition
in the vertical optical depth of the accreting gas.  The disk is quite
optically thick to both absorption and scattering, while the boundary
layer is still optically thick to scattering but optically thin to
absorption, even when the effects of scattering are included.  The
scattering optical depth also reaches a minimum of $\tau_s \sim 4$ in
the boundary layer, measured from the midplane to the surface.  The
radial optical depth through the boundary layer is of the same order
as the vertical optical depth.

When the infalling gas becomes optically thin to absorption, the gas
temperature rises dramatically as radiative cooling becomes
inefficient (Fig. 3).  (Note that the electron temperature shown in
the plots is always the midplane temperature; the electron temperature
near the disk surface may be quite a bit lower, depending on the
optical depth of the gas, and on the vertical dependence of the energy
dissipation.)  The effective temperature also peaks in this region at
$T_{eff} \sim 2 \times 10^7$ K; this is simply a measure of the flux
from the disk surface $F_V = \sigma T_{eff}^4$ and is not indicative
of the spectrum produced by the boundary layer gas.

The high temperature and radiation pressure causes the gas to expand
vertically, further lowering the density and the absorptive opacity.
The vertical expansion is quite dramatic, with the vertical scale
height going from less than 1\% of the stellar radius in the innermost
part of the disk to $\sim 25\%$ of the stellar radius in the boundary
layer (Fig. 3).  Nonetheless, the implied vertical velocity of the gas
$v_z \simeq v_R dH/dR < c_s$, so the gas stays in hydrostatic
equilibrium.

As the gas moves inward through the hot boundary layer, $\Omega$ drops
rather gradually, and the pressure increases again.  Note that almost
all of the pressure in the hot region is due to radiation, while in
the innermost disk gas pressure dominates (Fig. 4b).  The outward
pressure gradient provides support against gravity, compensating for
the sub-Keplerian rotation (Fig. 5).  Bulk viscous effects also
play a small but non-negligible role.  Note that the glitches in the
pressure gradient and bulk viscous terms at $R \simeq 1.038 \times
10^6 \cm$ in Fig. 5 are a numerical problem associated with a change
in grid resolution at that point.  At $R \simeq 1.04 \times 10^6 \cm$,
$\Omega$ is around 75-80\% of the Keplerian value, so about 60\% of
the support against gravity comes from rotation.  

Here the gas begins to pile up on the surface of the star, with $v_R$
decreasing quite rapidly (Fig. 3).  The gas makes a transition back to
an optically thick state, and the temperature drops rather abruptly.
$\Omega$ drops quite rapidly to the stellar rotation rate.  The height
of the disk also drops abruptly, so that the disk is again rather thin
as it reaches the stellar surface.  This seems somewhat
counterintuitive, and may be a consequence of using simple
one-dimensional disk equations to approximate a fundamentally
two-dimensional flow.  One might expect the gas to stay at a large
height above the midplane and fall onto the stellar surface in a wide
belt, rather than falling back toward the midplane.  Even if this is
the case, we expect the qualitative picture presented here to be
correct, since whether the local vertical scale height is 5\% or 50\%
of $R_*$, the gas must slow down and get much denser when it reaches
the stellar surface.

The innermost zone of the solution, inside $R \simeq 1.035 \times 10^6
\cm$, serves as the ``star'' in our calculation.  This portion of the
solution is calculated using the same disk equations which are used
everywhere else.  The gas settles slowly inward, maintaining the same
$\dot M$ as in the disk and boundary layer, with very low radial
velocity and very high density.  The energy balance is between inward
advection of the gas entropy and outward radial radiation flux.  The
gas is quite hot and the vertical (i.e. meridional) pressure scale
height is generally around 5--10\% of $R$.  Gas pressure again
dominates over radiation pressure here; this is difficult to see in
Fig. 4a due to the steep increase in both the total pressure and the
radiation pressure, but it can be seen clearly in Fig. 4b.  The
viscosity given by the $\alpha$ prescription is very high and so
$\Omega$ is nearly constant at the value set by the boundary
conditions, $\Omega \simeq \Omega_*$.

The mean intensity of radiation $u$ is several orders of magnitude
smaller than the blackbody mean intensity $B$ (calculated from the
electron temperature) in the hot, low-density region of the boundary
layer (Fig. 4c).  The temperature (and thus $B$) is high within this
low optical depth zone; however, $u$ decreases monotonically with $R$.
The photon and electron temperatures diverge in the same region
(Fig. 4d).  In the outer part of the boundary layer, the photon
temperature exceeds the electron temperature and {\it Comptonization
heats the electrons} (cf. Eq. 14 and the following discussion).
Inside $R \simeq 1.08 \times 10^6$ cm, the electron temperature
exceeds the photon temperature, so the photons cool the electrons due
to Comptonization.  Note that the ion temperature also diverges from
the electron temperature in the boundary layer region, reaching $\sim
10^9$ K.

The radiation flux is predominantly radial in the inner part of the
boundary layer (Fig. 4e) - this radiation carries away the energy
dissipated in the region where $\Omega$ drops rapidly.  This radiation
is scattered by the hot boundary layer gas and escapes from its
surface.  We have used the angle $\tan^{-1} (F_R/F_V)$ to indicate the
general direction of the radiation (Fig. 4f).  It should be noted that
$F_R$ is the radial flux at the disk midplane where $F_V$ is zero by
symmetry, while $F_V$ is the vertical flux at the disk surface where
$F_R$ is generally small.  (Since we do not explicitly solve the
vertical structure of the disk, $F_V$ and $F_R$ are assumed to
linearly increase and quadratically decrease with distance from the
midplane, respectively; see PN95 for details.)  Thus the ratio
$F_R/F_V$ does not give the true direction of the flux at any real
location in the disk, but we have used it as a convenient way to show
how the average direction of the flux changes with radius.

In the outer part of the boundary layer, the dominant terms in the
energy equation are Compton scattering and energy advection (Fig. 5).
Note that the energy terms in this figure are vertically integrated so
that they are in units of energy flux per unit midplane area of the
disk.  In the outer boundary layer, the gas is heated by scattering of
energetic photons from farther in.  Inside $R \simeq 1.08 \times 10^6$
cm, the situation reverses; here the photons are cooling the hot gas.
Thus Comptonization has the net effect of carrying energy outward in
the hot region, while advection carries energy inward.  Throughout
this low-density region, the viscous dissipation is rather small
compared to the Compton and advection terms, and the emitted radiation
(the $\rho \kappa_a (B-u)$ term in Eq. 16a) is negligible.  This
changes when the gas density increases near the stellar surface
(Fig. 5).  Here the viscous dissipation is large, and the Compton
scattering deposits some energy, but dwindles as the electron and
photon temperatures reach equilibrium.  The denser gas radiates the
dissipated energy away efficiently; most of the emitted photons travel
radially outward into the hot, low-density region and are inverse
Compton scattered to higher energies.

\subsection{Viscous Energy Transport}

The strong peak in the viscous dissipation at the inner edge of the
boundary layer comes from two sources.  The first is the kinetic
energy lost by the gas as $\Omega$ drops rapidly.  However, an even
larger source is the energy transported radially by viscosity.

Figure 5 shows the viscous transport $\Omega N = \dot M (\Omega^2 R^2
- \Omega j \Omega_K(R_*) R_*^2)$, which carries energy outward in the
disk and inward in the boundary layer.  Note that the viscous
transport is zero at $R \simeq 1.16 \times 10^6 \cm$, where $\Omega$
reaches its maximum value.  The local rate of energy deposition by
viscous transport depends on the radial gradient of the transport rate
$-d(\Omega N)/dR = -\dot M d (\Omega^2 R^2 - \Omega j \Omega_K(R_*)
R_*^2 )/dR$.  Where $\Omega N$ decreases with radius, as in the inner
boundary layer from $R \simeq 1.035 - 1.0365 \times 10^6 \cm$, or in
the disk outside $R \simeq 2.5 \times 10^6 \cm$, viscous transport
deposits energy.  Between $R \simeq 1.0365 \times 10^6 \cm$ and $R
\simeq 2.5 \times 10^6 \cm$, $\Omega N$ increases with radius, so
viscous transport removes energy.

The rate of energy gain or loss due to viscous transport is shown in
the bottom panels of Fig. 5, together with the local viscous
dissipation rate and the rate of loss of kinetic energy, all in units
of the local gravitational energy release per unit radius $(G M \dot M
/ R^2)$.  The most dramatic result is that the viscous transport is
almost fully responsible for producing the very sharp peak in the
dissipation rate in the dense inner boundary layer.  The kinetic
energy loss is also quite concentrated, reaching more than 100 times
the local gravitational release, but the viscous transport deposits
energy at a rate exceeding 600 times the local gravitational release.
The middle panel shows that over the entire outer boundary layer, the
gravitational potential and kinetic energy lost by the gas is not
dissipated locally, but instead is transported inward and dissipated
in a very narrow layer near the surface of the star, where the gas
density is much higher than in the outer boundary layer.  

The peak dissipation occurs in a zone where the radial optical depth
for the emitted photons to reach the hot, low-density outer boundary
layer is about 20, as shown in Fig. 6.  The vertical optical depth at
this radius is much larger than 20, so most of the radiation travels
radially outward.  Note that in this figure it appears that viscous
transport deposits much more energy in the dense region than it takes
away from the low-density region, but this is simply a consequence of
the fact that we have plotted the energy deposition as a function of
the radial optical depth.  In fact, the viscous transport term can
only redistribute the energy within the boundary layer, and as Fig. 5
shows, the region from which viscous transport removes energy is far
larger than the region where it deposits energy.  The fact that the
energy is dissipated in the dense, optically thick gas, rather than
in the rarefied gas of the outer boundary layer, makes a major
difference in the radiation spectrum of the boundary layer.

Far from the star, Fig. 5 shows that the energy deposited per unit
disk surface area by viscous transport of energy $(1/4 \pi) [1 - (3/2)
j R_*^{1/2} / R^{1/2}] \dot M \Omega_K^2$ increases the viscous
dissipation rate to $Q^+ = (3/8 \pi)(1 - j R_*^{1/2} / R^{1/2}) \dot M
\Omega_K^2$.  At very large radii, this is three times larger than the
$(1/8 \pi) \dot M \Omega_K^2$ expected from the change in the
gravitational potential energy and kinetic energy of the gas.  The
energy deposited by viscous transport is positive for $R > (9/4) j^2
R_*$; in the solution shown in Fig. 5, $j \simeq 1.064$, so the
viscous transport deposits energy for $R \ga 2.55 R_*$ and removes
energy for $R \la 2.55 R_*$.

\subsection{Mass Accretion Rate}

Figure 7 shows the effects of varying $\dot M$.  The solutions cover
the range $\dot M = 10^{-8} - 10^{-10} \msyr$; the corresponding
accretion luminosities are $L_{acc} = G M \dot M / R_* = 1.17 \times
10^{37} (\dot M / 10^{-9} \msyr) \ergs$, part of which should be
emitted by the boundary layer and part by the disk.  The most apparent
effect of increasing $\dot M$ is the dramatic radial expansion of the
boundary layer region.  The location of the outer edge of the boundary
layer is clear in the figures; $\Omega$ deviates from $\Omega_K$ and
decreases as the gas falls inward, while all the other variables show
a sudden increase (in temperature, radial velocity and disk scale
height) or decrease (in density and optical depth).  At $10^{-10}
\msyr$ the radial extent of the region is only about 10\% of the
stellar radius, whereas at $10^{-8} \msyr$ it extends to $R \simeq 2
R_*$.  The dramatic increase in the boundary layer width for high
$\dot M$ occurs because $10^{-8} \msyr$ is very close to the Eddington
limit, as discussed further below.  

The profile of $\Omega$ in the boundary layer also changes as a
function of $\dot M$.  As discussed above, $\Omega$ drops more
gradually in the region where the density is lower, and then quite
rapidly in the denser region near the stellar surface.  As $\dot M$
increases, more of the drop in $\Omega$ occurs in the low-density
region: at $\dot M = 10^{-10} \msyr$, more than 90\% of the drop in
$\Omega$ is in the dense region, while at $10^{-8} \msyr$ less than
20\% is in the dense region.  This, together with the changing width
of the low-density region, means that the energy dissipation is
concentrated in a far smaller volume at low than at high $\dot M$.

The optical depth of the boundary layer increases steadily with
increasing $\dot M$.  It is marginally optically thick to scattering
at $10^{-10} \msyr$, but $\tau_s \ga 100$ for all radii at $10^{-8}
\msyr$.  Nonetheless, the free-free opacity is so low that the
boundary layer remains optically thin to absorption at all accretion
rates, even when the effects of multiple scattering are taken into
account.  Even at $10^{-8} \msyr$, the effective optical depth $\tau_*
= (\tau_s \tau_a)^{1/2} \simeq 0.001$.

The temperature profile of the boundary layer changes substantially as
a function of $\dot M$.  Solutions for all values of $\dot M$ show a
jump in $T$ in the low-density region, and over most of the region,
the higher values of $\dot M$ produce higher temperatures, ranging
from $\sim 1.5 - 5.5 \times 10^8$ K as $\dot M$ increases.  However,
the lower values of $\dot M$ produce double-peaked temperature
profiles, and the inner peak reaches higher temperatures as $\dot M$
decreases, reaching $T \sim 8 \times 10^8$ K at $\dot M = 10^{-10}
\msyr$.  The difference between the ion and electron temperatures is
also sensitive to $\dot M$; at $10^{-10} \msyr$, $T_{i,max} \sim 2
\times 10^{10} {\rm K} \sim 25 T_{e,max}$, while at $10^{-8} \msyr$,
the Coulomb coupling is much more efficient and there is practically
no difference between $T_i$ and $T_e$.

\subsubsection{The Ratio of Disk and Boundary Layer Luminosities}

One important effect of the radial expansion of the boundary layer is
to change the boundary layer and disk luminosities.  As we discussed
in \S 2.3, viscous energy transport cannot carry energy across the
maximum in $\Omega$ which occurs at the outer edge of the boundary
layer.  Thus, the luminosity of the disk must equal the change in the
gravitational potential and kinetic energy of the gas within the disk.
If the boundary layer is very small, then the disk and boundary layer
luminosities are about the same.  Half of the total accretion
luminosity is released in the disk, and the other half goes into
rotational kinetic energy, which is released in the boundary layer.
As the transition between the disk and boundary layer moves to radii
significantly larger than $R_*$, the amount of energy released in the
disk decreases, and that energy is radiated by the boundary layer
instead.  For instance, if the transition is at $R = 2 R_*$, only 1/4
of the total accretion luminosity is released in the disk, and 3/4 in
the boundary layer (1/4 from the kinetic energy of the gas at $R = 2
R_*$, and 1/2 from the gravitational energy released as the gas moves
from $2 R_*$ to $R_*$).  This would make the ratio of boundary layer
to disk luminosity 3:1 instead of the usual 1:1 (for the Newtonian
case; in the Schwarzschild metric the ratio is about 2:1 (Sunyaev \&
Shakura 1986)).  This could have important effects on the overall
radiation spectrum.

\subsubsection{Effect of Flux from the Neutron Star}

Note that we have assumed in our calculations that the radiative flux
across $R_*$ is very small.  The accreted hydrogen should burn on the
surface of the star and release a luminosity $\sim 0.007 \dot M c^2
\simeq 4 \times 10^{35} (\dot M / 10^{-9} \msyr) \ergs$.  If radiated
evenly from the whole surface of the star, this produces a flux $\sim
3.2 \times 10^{22} (\dot M / 10^{-9} \msyr) \ecps$.  We have carried
out calculations in which the radiative flux at the inner boundary is
set to this value.  The resulting solutions are almost identical to
those with no flux at the boundary.  The reason for this is clear from
Fig. 8; the radial flux at the inner edge of the boundary layer
reaches $\sim 0.1 - 1.4 \times 10^{25} \ecps$ for $\dot M =
10^{-10}-10^{-8} \msyr$, respectively, a significant fraction of the
Eddington flux.  The flux from nuclear burning on the stellar surface
would have to reach a significant fraction of this value before it had
a major effect on the boundary layer structure.  Such fluxes are
probably reached during X-ray bursts.

\subsection{Accretion Rates Near the Eddington Limit}

The radial extent of the boundary layer becomes much larger at $\dot M
= 10^{-8} \msyr$.  This accretion rate is close to the Eddington
limiting rate, which for a spherical geometry is $\dot M_{Edd} = 4 \pi
c R_* / \kappa_s \simeq 1.73 \times 10^{-8} \msyr$ for $R_* = 10^6
\cm$.  In a disk geometry, the crucial quantities are the radiation
flux in the radial direction $F_R$ and the Eddington limiting flux
$F_{Edd} = (G M / R^2) c / \kappa_s$; these are shown in Fig. 8.  The
radiation provides an increasing fraction of the support against
gravity as the gas moves inward through the boundary layer.  This is
reflected in the gradual decrease of the radial velocity $v_R$ as the
gas flows through the boundary layer in the $10^{-8} \msyr$ solution.
Note that the radial velocity profile is rather different from the
solutions for lower values of $\dot M$ which are not so close to the
Eddington limit (Fig. 7).

Even though the radial flux in the $\dot M = 10^{-8} \msyr$ solution
is very close to the local Eddington value, we are able to find
solutions for larger values of $\dot M$.  These solutions are shown in
Fig. 9.  The radial expansion of the boundary layer is dramatic: at
$\dot M = 10^{-7.85} \msyr \simeq 1.4 \times 10^{-8} \msyr$, the outer
edge of the boundary layer is at $R \simeq 3 \times 10^6 \cm$.  Thus
the radial extent of the boundary layer doubles for a 40\% increase in
$\dot M$.  Note that $1.4 \times 10^{-8} \msyr$ is still less than the
Eddington limit for spherical accretion, and of course part of the
luminosity is released in the disk, so the luminosity released in the
boundary layer is well below the spherical Eddington limit.  On the
other hand, the disk covers only a fraction of the stellar surface, so
the outgoing radial flux within the disk comes very close to the
Eddington limiting flux.

Overall, the transition from disk to boundary layer is much less
abrupt than at lower accretion rates.  For example, $\Omega$ gradually
deviates from Keplerian, so that $\Omega / \Omega_K$ drops steadily,
but $\Omega$ itself continues to increase and reaches a maximum
farther in.  At lower accretion rates, $\Omega$ is fairly strictly
Keplerian within the disk, reaches its maximum at the boundary layer
transition, and then decreases within the boundary layer.  The
different $\Omega$ profiles have consequences for the amount of
angular momentum accreted by the star, as discussed below.  Note that
the temperature profile is quite smooth, in marked contrast to lower
accretion rates.  Also, the minimum in temperature at the transition
disappears; this is due to heating of the inner disk by the boundary
layer radiation.  The height of the disk is a large fraction of the
radius, $H/R \sim 0.6 - 0.8$, and $H/R$ stays fairly constant over
much of the radial extent of the boundary layer.  The boundary layer
stays optically thick to scattering but optically thin to absorption.

\subsection{Viscosity Dependence}

The radial velocity of the accreting gas is approximately proportional
to the viscosity coefficient, since this determines the rate at which
angular momentum is removed from the gas.  As a result, for a given
mass accretion rate, the disk surface density is inversely
proportional to viscosity.  Thus, changing the value of $\alpha$
changes the optical depth of the disk.

We illustrate the effects of $\alpha$ in Fig. 10, which shows solutions
with $\alpha=0.01$ and $\dot M = 10^{-8} - 10^{-10} \msyr$.  In most
respects this sequence looks similar to that shown in Fig. 7 above for
the same accretion rates with $\alpha=0.1$.  The width of the boundary
layer is quite similar in the two sequences, especially at high $\dot
M$; at $\dot M = 10^{-10} \msyr$ the $\alpha = 0.01$ boundary layer is
about half as wide as it is for $\alpha = 0.1$.  The disk height is
also quite similar - at moderate accretion rates it is slightly larger
in the $\alpha = 0.01$ solutions.  The effective temperatures are also
quite similar, as one would expect, since independent of $\alpha$, the
same amount of energy is being released in an area of approximately
the same size.  Also as expected, the $\alpha = 0.01$ solutions have
lower $v_R$ and higher $\rho$ and $\tau_s$ than the $\alpha=0.1$
solutions, in all cases by a factor of $\sim 10$.

The main qualitative difference between the two sets of solutions is
in the gas temperature in the boundary layer, which is always lower
for the $\alpha = 0.01$ solutions.  The boundary layer becomes
marginally optically thick for $\alpha=0.01$ at $\dot M = 10^{-8}
\msyr$, with $\tau_* \sim 1$; as a result, the temperature does not
jump dramatically at the outer edge of the boundary layer, but
increases steadily as the gas approaches the star.  The
$\alpha=0.01$ solutions do not show the strongly double-peaked
temperature profile seen at low $\dot M$ for $\alpha=0.1$.  Also, even
at $10^{-10} \msyr$, the ion and electron temperatures are very close
in the $\alpha=0.01$ solutions.

In all the solutions so far, we have used the ``standard'' viscosity
prescription, which takes the turbulent length scale $l_{turb} =
(H^{-2} + H_R^{-2})^{-1/2}$, which is approximately equal to the
lesser of the vertical and radial pressure scale heights $H$ and $H_r$
in the disk.  If we compare $H$ and $H_r$ with the length scale for
subsonic turbulence $H_s$, we find $H_s < H_r, H$ in much of the
boundary layer, as shown in Fig. 11a for the $\dot M = 10^{-9} \msyr$,
$\alpha = 0.1$ solution.  This means that the azimuthal velocity
difference across the adopted turbulent length scale is supersonic.
However, note that since $\alpha = 0.1$, the viscosity coefficient
$\nu = \alpha c_s l_{turb}$ is less than the product of $c_s$ and the
subsonic length scale $H_s$; in other words, over a length scale
$\alpha l_{turb}$, the azimuthal velocity difference is still
subsonic.

We have calculated a set of solutions using the ``subsonic''
viscosity prescription outlined in \S 2; here $l_{turb} = (H^{-2} +
H_R^{-2} + H_s^{-2})^{-1/2} \sim \min(H,H_r,H_s)$ (Fig. 11b).  The
results are shown in Fig. 12.  At high $\dot M$, the subsonic
prescription changes the solution very little, but at lower $\dot M$,
it substantially decreases the viscosity in the boundary layer.  As a
result, the lower $\dot M$ solutions with $\alpha=0.1$ and subsonic
viscosity look very much like the solutions with $\alpha=0.01$ and
standard viscosity.  In the boundary layer, they have lower
temperatures ($T_e \sim 10^8$ K) and radial velocities and higher
densities and optical depths than the standard viscosity solutions.

Solutions with subsonic viscosity can also reach higher temperatures
of a few $\times 10^8$ K when higher values of $\alpha$ are used, as
shown in Fig. 13.  In the upper panels, we compare solutions with
$\dot M = 10^{-9} \msyr$ and subsonic viscosity for $\alpha=0.1$ and
0.2 to a solution with standard viscosity and $\alpha=0.1$.  The
solutions with subsonic viscosity reach higher boundary layer electron
temperatures and lower scattering optical depths as $\alpha$
increases, and more closely resemble the solution with regular
viscosity and $\alpha=0.1$.  Similarly, in the lower panels we show
solutions for $\dot M = 10^{-10} \msyr$ and subsonic viscosity with
$\alpha= 0.1$ and 0.316; the electron temperature and scattering
optical depth for the $\alpha=0.316$ solution are closer to the
solution with standard viscosity and $\alpha=0.1$.

\subsection{Rotation and Accretion Spinup of the Star}

The coherent kHz oscillations recently discovered during X-ray bursts
from LMXBs strongly suggest that many of these contain neutron stars
which are rotating at speeds of $\sim 300$ Hz (or possibly $\sim 600$
Hz) (Strohmayer \et 1996).  Rotation of the accreting star should have
an important impact on the boundary layer region.  The energy
dissipated in the boundary layer is proportional to $(1 -
\Omega_*/\Omega_K)^2$, so it decreases rather rapidly as the star
spins up.

In Fig. 14 we show the structure of the boundary layer for five
different stellar rotation frequencies $f_* = 0, 318, 637, 1273,
1910$ Hz (corresponding to $\Omega_* = 2 \pi f_* = 0, 2000, 4000,
8000, 12000 \pers$).  Based on current observations, the lower values
are the most relevant; however, the effects of rapid rotation on the
boundary layer and spinup of the star are interesting enough that we
have included solutions for rotation speeds extending up to nearly
breakup.  All of the solutions are for $\dot M = 10^{-9} \msyr$ and
$\alpha=0.1$.  The width and effective temperature of the boundary
layer decrease steadily as the star spins up.  The low-density region
becomes cooler and denser and the optical depth increases.  At
$\Omega_* = 12000 \pers$ the boundary layer is becoming marginally
optically thick.  It should be noted that rapid rotation will flatten
the star and increase its equatorial radius, and we have not included
this in our calculations; we do not expect that it will qualitatively
change our results.

In our calculations we specify the angular momentum accretion rate
$\dot J$.  The thin disk equations assume that the angular momentum
accretion rate onto the star is $\dot J_{thin} = \dot M \Omega_K (R_*)
R_*^2$.  The angular momentum balance then gives a factor $1 -
(R_*/R)^{1/2}$, which appears in the thin disk solutions and produces
a maximum in the effective temperature at $R = (49/36) R_*$ (Shakura
\& Sunyaev 1973).  Inside this radius the effective temperature drops
again, and formally goes to zero at $R = R_*$.

In our solutions $\dot J \neq \dot J_{thin}$.  In general, $\Omega$
reaches a maximum at some radius $R_{max}$ which is near the
transition from the disk to the boundary layer, and $\dot J$ is set by
the angular momentum of the gas at $R_{max}$, $\dot J = \dot M
\Omega_{max} R_{max}^2$.  We specify $\dot J$ using the parameter $j
\equiv \dot J / \dot J_{thin}$, so $j = \Omega_{max} R_{max}^2 /
\Omega_K (R_*) R_*^2$.  If $\Omega \simeq \Omega_K$ in the disk and
$\Omega_{max} \simeq \Omega_K(R_{max})$, then the angular momentum
equation can be solved in the usual way, except that the factor in the
disk equations becomes $1 - (R_{max}/R)^{1/2} \simeq 1 - j
(R_*/R)^{1/2}$, i.e. $j \simeq (R_{max}/R_*)^{1/2}$.  The maximum in
$T_{eff}$ occurs at $R \simeq (49/36) R_{max} \simeq (49/36) j^2 R_*$.
In the solution shown in Figs. 3--5, $j \simeq 1.064$, so $T_{eff}$
should reach a maximum at $R \simeq 1.54 R_*$.

The condition that the disk outside of $R_{max}$ is close to Keplerian
holds true in most of our solutions, with the exception being the
solutions at high $\dot M$.  There $\Omega$ is substantially
sub-Keplerian when it reaches a maximum, and as a result the value of
$j$ can be quite a bit less than $(R_{max}/R_*)^{1/2}$.  For example,
our solution with $\dot M = 10^{-8} \msyr$ shown in Fig. 7 has $j
\simeq 1.155$, even though $R_{max} \simeq 1.84 R_*$, because
$\Omega_{max}$ is only $\simeq 0.85$ of $\Omega_K(R_{max})$.

Note that although we specify the value of $j$, it is fairly tightly
constrained.  If we increase $j$, then $R_{max}$ and the whole
solution move outward in radius.  We have selected values of $j$ for
which the inner edge of the boundary layer, where the gas piles up on
the star, is located at $R \simeq 1.035 \times 10^6 \cm$.  This value
is selected arbitrarily; it leaves a small zone between $R = 10^6 \cm$
and $R \simeq 1.035 \times 10^6 \cm$ where the accreting gas is piling
up on the star.  By choosing slightly smaller or larger values of $j$,
we could have located the boundary layer at a slightly smaller or
larger radius.

As an accreting star spins up near breakup, the rate of angular
momentum accretion drops rapidly and becomes negative (Popham \&
Narayan 1991; Paczynski 1991).  We have found solutions with $j < 1$,
and we show a solution with $j = 0$ at $\Omega_* = 1.31 \times 10^4
\pers$ as a dashed line in Fig. 14.  Here the boundary layer is
absent, and the accretion flow joins the star smoothly.  For larger
values of $\Omega_*$, $j$ becomes negative; thus this value of
$\Omega$ should be the maximum reached by the star for this choice of
parameters.

\section{Discussion}

\subsection{Energetics of the Boundary Layer}

We have found that boundary layers around disk-accreting neutron stars
will be hot, low in density, optically thin to absorption, and both
radially and vertically extended.  We begin by discussing some of the
processes which are important in producing this boundary layer
structure.  We focus particularly on the important role played by
radial energy transfer.

We find that the accreting gas makes a fairly abrupt transition from
the disk to the boundary layer.  The disk is thin, relatively cool and
dense, and optically thick to absorption, while the boundary layer is
geometrically thick, hot and rarefied, and optically thin to
absorption for most choices of parameters.  There is a similar abrupt
transition in reverse when the hot gas nears the stellar surface.
These transitions are related to the thermal instability of the hot,
low-density boundary layer gas discussed by King \& Lasota (1987), in
which the gas cannot efficiently radiate away the dissipated energy.
However, the situation is more complicated than the one they
envisioned, since they confined their analysis to local heating and
cooling by dissipation and radiation, in a gas-pressure-dominated disk
(corresponding to lower accretion rates and luminosities than those
considered here).  In our solutions radiation pressure is dominant and
the energy balance is dominated not by local dissipation and radiation,
but instead by Comptonization and advection.

In our solutions the gas reaches temperatures of a $\few \times 10^8$
K, and nearly reaches $10^9$ K in one case.  One might expect the
large energy release to make the boundary layer gas even hotter than
this; if all the accretion energy were to go into heating the gas, it
would reach the virial temperature of a $\few \times 10^{11}$ K.
However, the presence of Compton cooling keeps the electron
temperature from rising above $\sim 10^9$ K.  Solutions computed
without Compton cooling reached very low optical depths and high
temperatures in excess of $10^9$ K even for very small values of
$\alpha = 10^{-4} - 10^{-3}$. For $\alpha=0.1$ these temperatures
would be much higher, and would presumably approach the virial
temperature.  This illustrates the dominant role played by
Comptonization in transferring energy from the gas to the photons and
cooling the boundary layer region.

We have solved separate energy equations for the electrons and ions,
allowing for the possibility of a two-temperature plasma.  In most of
our solutions, the ion and electron temperatures are essentially the
same at all radii; however, in solutions with $\dot M \la 10^{-9}
\msyr$ and $\alpha \ga 0.1$, the ion temperature is significantly
higher than the electron temperature in the boundary layer region.
The ion temperature increases rapidly as $\dot M$ decreases, reaching
$\sim 2 \times 10^{10}$ K at $\dot M = 10^{-10} \msyr$, $\alpha=0.1$.
This increased ion temperature produces an increase in the gas
pressure, and together with the smaller luminosity, this will lead to
gas pressure becoming the dominant source of pressure at low $\dot M$.

Radial transport of energy by advection also plays an important role
in our solutions.  This is not surprising, since NP93 found that
advection plays an important role in hot CV boundary layers, which led
to the ``rediscovery'' of advection-dominated accretion by Narayan \&
Yi (1994).  The importance of advection can be seen simply by noting
that the disk is not geometrically thin in the boundary layer region,
which means that the energy density is significant compared to
gravity.  However, the flow in the boundary layer differs in some
respects from the standard advection-dominated accretion flow (ADAF),
which is heated by viscous dissipation and cooled (very inefficiently)
by Compton cooling, so that essentially all the energy goes into
heating the gas.  In the boundary layer, as discussed in \S 3, the
energy balance is between Compton heating (or cooling) and advection.
In the outer boundary layer, the energetic photons contribute far more
energy than viscous dissipation, and the gas cannot radiate this
energy away, so it heats up and expands and the energy is carried
inward.  In the inner boundary layer, this situation is reversed.
Thus, even though the situation differs somewhat from a standard ADAF,
advection is nonetheless an important energy term throughout the hot
region.

One of the main predictions of the mechanics of the boundary layer is
that viscous transport carries energy inward, toward the dense,
optically thick region at the surface of the star.  The situation is
very similar to that in the standard disk (Shakura \& Sunyaev 1973),
where viscous transport carries energy away from the central regions
of the disk and delivers it to larger radii, increasing their energy
release by a factor of three compared to the simple estimate based on
the the change in the gravitational and rotational energy of the
infalling gas.  In the boundary layer, where $d \Omega/dr$ has a
different sign, the viscous torque transports energy toward the star
and this leads to important consequences.

This can be seen in Figure 5, which shows the energy release due to
viscous dissipation, the viscous transport of energy, and the change
in kinetic energy.  In the dense region in the innermost part of the
boundary layer, the rotational velocity of the gas is much smaller
than Keplerian.  If we consider the deepest layer, where the radiation
flux is still much smaller than the local critical Eddington flux, we
see immediately that the density of the matter must be very high,
because it is not supported strongly by centrifugal forces or by the
radiation pressure gradient $dP_{rad}/dR \propto F_R$.  A large
fraction of the energy is dissipated in, and subsequently radiated
away from, this high-density region.  Much of the radiation travels
radially outward into the hot, low-density region, providing the seed
photons for Comptonization.  In the solution shown in Figs. 3--5, the
density at the point where the dissipation peaks is nearly $1 \gcm$,
and the radial scattering optical depth $\tau_s = \int \rho \kappa_s
dR$ between that radius and the radius where the temperature peaks is
$\sim 20$, as shown in Fig. 6.  It is crucial for the
formation of the spectrum that viscous transport and advection carry
the bulk of the energy from the low-density, rapidly rotating outer
boundary layer to the dense, slowly rotating region where the gas
reaches the stellar surface.

\subsection{The Eddington Flux Limit}

The Eddington flux limit, where the outward radiation pressure
gradient balances gravity, plays an important role in our solutions.
For spherical accretion, $\dot M$ is limited to $\dot M_{Edd} = 4 \pi
c R_* / \kappa_s \simeq 1.73 \times 10^{-8} \msyr$ for $R_* = 10^6
\cm$.  In a disk geometry, where the luminosity is not emitted
isotropically, the crucial quantity is the critical Eddington flux
which balances the local gravitational force.  The radial flux through
the disk is limited to the local Eddington value $F_{Edd} = (G M /
R^2) c/\kappa_s$.  In the vertical direction, the downward component
of gravity increases with distance $z$ from the midplane as $\sim
z/R$.  Thus at the disk surface $z \sim H$, the vertical flux is
limited to $F_{Edd,V} = (H/R) (G M / R^2) c/\kappa_s$.

For high values of $\dot M$, the radial and vertical radiation fluxes
approach their respective Eddington limiting fluxes.  In Fig. 8 we
showed that for $\dot M = 10^{-8} \msyr$, the radial flux comes very
near the Eddington value, but only over a small range of radius, and
drops steadily as the radiation moves outward through the boundary
layer.  This reflects the fact that more of the radial support against
gravity comes from the centrifugal force and less from radiation
pressure as one moves outward through the boundary layer.  The
vertical flux, on the other hand, stays within 2\% of the Eddington
value throughout the entire boundary layer.  Even at $\dot M = 10^{-9}
\msyr$, the vertical flux stays between 80\% and 90\% of $F_{Edd,V}$.
This close correspondence between $F_V$ and $F_{Edd,V}$ means that the
radiation flux from the boundary layer has a radial profile varying as
$H/R^3$.  Physically, it means that at high $\dot M$ the boundary
layer is radiating as much flux as it can.

In order to increase $\dot M$, the boundary layer must radiate more
energy, and in order to do this it must expand either radially or
vertically, or both.  By expanding radially, it increases the surface
area through which the energy can escape, while vertical expansion
increases the gravity and the limiting flux.  In the outer portion of
the expanded boundary layer, both the radiation flux and centrifugal
forces are important in supporting matter against gravitational
attraction to the neutron star.  The rotational velocity decreases
significantly in this region as the gas flows inward, and is replaced
by radiation pressure support; however, the energy release due to
viscous dissipation is very small, since as we have seen, viscous
transport carries most of the energy to be dissipated farther in.
Thus this region serves mostly to radiate the energy which is
dissipated farther in.

We have shown that the boundary layer is both radially and vertically
extended, with the radial width reaching $\sim R_*$ and $H/R \simeq
0.8$ at $\dot M \sim 10^{-8} \msyr$ (Fig. 7).  If we continue to
increase $\dot M$, the boundary layer continues to expand radially and
increase its emitting area (Fig. 9).  The radial expansion is quite
rapid, with the boundary layer size doubling for a 40\% increase in
$\dot M$.  By increasing the emitting area, the radial expansion makes
it possible for the boundary layer to radiate away the very high
luminosities associated with these high values of $\dot M$, while
keeping the vertical flux below the Eddington limit.

This illustrates an important difference between disk accretion and
spherical accretion.  In spherical accretion, the Eddington limit is
global, in the sense that any luminosity produced inside a given
radius must contribute to the total outward flux at that radius.  Thus
the total luminosity of the system is constrained by the Eddington
limit.  In a disk system, the Eddington limit is local; the local
radial and vertical fluxes cannot exceed the local gravity.  However,
there is no global limit on the luminosity, since it can be radiated
away through the surface area of the boundary layer, and as we have
seen, the boundary layer can expand to radiate additional luminosity
as needed.

\subsection{Radiation Spectrum}

The structure of the boundary layer described above permits simple
modeling of the formation of the radiation spectrum which leaves the
boundary layer.  We plan to calculate detailed spectra based on our
solutions in a future paper; however, we can make some general
statements.  It is clear that Compton scattering will play a dominant
role in the formation of the X-ray spectrum of the radiation which
travels through the hot, low-density region.  The seed photons which
are Compton-scattered are primarily emitted from the denser gas which
piles up at the inside edge of the hot region.  The Comptonized
spectrum will be characterized by $T_e$ and $\tau_s$, and especially
by the Compton $y$-parameter $y = (T/T_{com}) max(\tau_s, \tau_s^2)$.

These parameters vary with radius in our solutions, and of course
$T_e$ and $\rho$ also vary with vertical position in the disk, and
this variation is not included directly in our solutions.  Also,
$\tau_s = \kappa_s \rho H$ in the vertical optical depth for our
solutions, but the photons are not travelling only vertically, but
also radially and azimuthally, and being scattered in all directions,
so the path traveled by a given photon may be much longer than $H$.
Therefore, although we can get a rough sense of what the X-ray
spectrum will look like based on the characteristic values of $T_e$
and $\tau_s$ for a given solution, the formation of the real spectrum
will be considerably more complex.

In the optically thick region, where Thompson scattering nevertheless
strongly dominates the opacity, free-free processes easily produce
many photons at low frequencies where they ($K_\nu \sim \nu^{-2}$) are
more effective than Comptonization.  Comptonization increases the
energy of low frequency photons due to the Doppler effect and leads to
the diffusion of the photons towards higher frequencies.  At some
frequency $x_0 = {{h\nu_0\over kT_e}}$ the rate at which
Comptonization can take photons and bring them to higher frequencies
equals the rate of photon absorption due to bremsstrahlung.
Practically every photon born with a frequency higher than $x_0$ will
be transported toward the frequency $x=3$ due to the Comptonization
process.  This picture is very similar to the processes occurring in
the early universe and is described in detail by Illarionov \& Sunyaev
(1975).  Comptonization is very effective because the parameter $y =
{{kT_e}\over{m_ec^2}} \tau_s^2 \gg 1$.  Under these circumstances a
Wien-type spectrum must be formed when $x_0$ is very small and $y \ga
1$ but not very high.  At higher $\tau_s$ and $y$ a Bose-Einstein
spectrum is formed and the formation of the black body spectrum is
possible inside very deep regions.

Radiation with this spectrum diffuses out from the dense regions
toward more and more rarified regions, where the production of new
soft photons becomes more and more difficult due to the low density.
Comptonization continues to dominate, and therefore the spectrum tends
to be close to a Wien spectrum, with a strong increase of intensity at
low frequencies where the spectrum becomes a Rayleigh-Jeans spectrum.
The temperature of electrons in the outer regions is determined by the
radiation: high energy photons heat electrons due to the recoil
effect, and low energy photons cool them down.  (See Levich \& Sunyaev
1971).  This process is illustrated by Fig. 5. We see that
Comptonization takes energy from the plasma in the inner part of the
boundary layer and gives energy back to the plasma in the outer part,
heating the electrons.  As a result, we are producing a
quasi-Wien-type radiation spectrum with strong low frequency excess,
and radiation with such a spectrum escapes from every point of the
surface of the boundary layer.  Different regions of the boundary
layer have a range of temperatures, depending on the distance from
both the stellar surface and from the midplane of the disk.  Therefore
we will observe a spectrum which is a sum of Wien spectra with
different temperatures.

The picture described above follows the results of extended
calculations by Grebenev \& Sunyaev (2000) for the spreading layer
model (Inogamov \& Sunyaev 1999) of the surface of the neutron star.
It is interesting to note that this ``spreading layer'' picture of the
boundary layer, in which the gas loses angular momentum as it spreads
over the stellar surface, gives rather similar results to the ones
presented here.  In both cases, the size of the layer increases as the
mass accretion rate and luminosity increase.  The ``spreading layer''
has a meridional extent of about 0.45 km at $L/L_E = 0.01$ (which
should correspond approximately to our $\dot M = 10^{-10} \msyr$
solutions), increasing to $\sim 17$ km at $L/L_E = 0.8$.  In our
solutions the radial extent of the boundary layer is $\sim 1, 0.5$
km for $\dot M = 10^{-10} \msyr$ and $\alpha=0.1,0.01$, respectively,
and $\sim 10$ km for $\dot M = 10^{-8} \msyr$.  In addition, the
optical depths and $y$-parameters are quite similar in the two types
of solutions.  Inogamov \& Sunyaev (1999) found $\tau_s \sim 3$ at
$L/L_E = 0.01$ and $\tau_s \sim 1000, y \sim 10^5$ at $L/L_E = 0.8$,
quite similar to the values presented above.  Thus the two treatments
should result in rather similar spectra.  Most importantly, in both
cases the majority of the energy release occurs in a dense region
which is covered by a levitating low density region where the final
spectrum is formed.  The agreement is remarkable considering that the
accretion flow is treated quite differently in the two approaches.

The question of which approach is the correct one is difficult to
answer at present.  The main difference between them is the assumed
geometry of the boundary layer region.  The ``spreading layer''
treatment assumes that the disk material enters the spreading layer at
nearly Keplerian rotation velocity and small disk height; i.e., very
little angular momentum is lost in the disk and there is no disk
boundary layer of the type calculated in the present paper.
Conversely, the approach taken here assumes that the drop from
Keplerian to the stellar angular velocity takes place during the
radial inflow of the gas, rather than during the spreading of the
accreted gas over the stellar surface.  A multi-dimensional treatment
will be required to distinguish between these two possibilities.  The
treatment of viscosity in the disk and surface layer may have an
important impact on the results.  Note that the viscosity prescription
adopted by Inogamov \& Sunyaev (1999) is also quite different from the
one used here; the viscosity decreases as the gas approaches the
neutron star surface in analogy with the behavior of fluid near a wall
in laboratory experiments.  Yet despite these differences their
results are rather similar to ours.

\subsection{Comparison to Observed Spectra of LMXBs}

Most observed LMXBs have been in one of two spectral states: a low
state characterized by low luminosity and a hard, power-law spectrum,
or a high state characterized by high luminosity and a softer
spectrum.  For example, the four LMXBs recently observed by Barret \et
(1999) included three in the low state and one in the high state.  The
three low state sources 1E1724-3045, GS1826-238, and SLX1735-269, all
had $1-200 \keV$ luminosities of $\sim 1 - 1.5 \times 10^{37} \ergs$,
and their spectra were fitted by thermal Comptonization by gas with
electron temperatures of $\sim 25-30$ \keV and optical depths of a
few.  The high state source KS1731-260 had $L (1-200 \keV) \simeq 8-9
\times 10^{37} \ergs$, and was fitted by a much softer Comptonized
spectrum with $T_e \sim 2.6-2.8$ keV and $\tau \sim 10$.

Our $\dot M = 10^{-10}$ solution with $\alpha=0.1$ and regular
viscosity (shown in Fig. 7) has $T_e$ in the hot boundary layer
varying from $\sim 10^8 - 10^9$ K and $\tau_s \sim 0.5 - 1$, giving a
Compton $y$-parameter of less than 1.  The large variations in $T_e$
and $\tau_s$ make it difficult to predict what the spectrum will look
like, but probably it will have a general power-law shape, and may
extend to rather high energies due to the very high $T_e$ in the
hottest part of the boundary layer.

At moderate $\dot M \sim 10^{-9} \msyr$, $T_e \sim 2-3 \times 10^8$ K
and both $\tau_s$ and $y \sim$ a few.  Unsaturated Comptonization in
the hot boundary layer should produce a power-law spectrum with a
cutoff at $\sim 20-30$ keV.  This solution has a total luminosity of
$\sim 10^{37} \ergs$, which corresponds approximately to that of the
low-state LMXBs oberved by Barret \et (1999), and the values of $T_e$
and $\tau_s$ agree well with those from their fits.

At high $\dot M$, $\tau_s$ is in the hundreds (for $\alpha=0.1$) or
thousands (for $\alpha=0.01$) and $y \sim 10^5$.  The optical depth
for combined absorption and scattering is $\tau_* \sim 0.1$ for
$\alpha=0.01$ but only $\tau_* \sim 0.001$ for $\alpha=0.1$.  Saturated
Comptonization will then produce a Wien spectrum which peaks at $\sim
3 k T$ (Illarionov \& Sunyaev 1975).  $T$ varies from $\sim 10^8$ K
for $\alpha=0.01$ to $\sim 3 \times 10^8$ K for $\alpha=0.1$.  This
will produce spectra with $kT \sim 10-30$ keV.  However, these are the
spectra that will be produced by the gas near the disk midplane, but
these solutions have such large values of $\tau_s$ and $y$ that we
expect a substantial temperature gradient, with the gas near the
surface much cooler than at the midplane.  Thus the spectrum will
strongly resemble a blackbody spectrum with a characteristic
temperature which is close to the effective temperature of the gas
rather than the midplane temperature, i.e. around 1.5--2 keV.
Blackbody fits to luminous LMXBs (e.g. Mitsuda \et 1984) give $kT \sim
2$ keV, while White \et (1986) used a $1-2$ keV blackbody plus an
unsaturated Comptonized component with a cutoff at $3-8$ keV.

Overall, it appears that our solutions should fit the spectral data
reasonably well.  The variation of boundary layer temperature with
$\dot M$ depends on $\alpha$; for low $\dot M$, the boundary layer is
much hotter for $\alpha=0.1$ than for $\alpha=0.01$.  Since
low-luminosity LMXBs generally are observed to have rather hard
power-law tails which are fit using electron temperatures of $\sim 30
\keV$, it seems that our models will agree with the observations
better if a large value of $\alpha$ is used.

X-ray spectra of LMXBs frequently show evidence for iron line emission
at about 6.4 keV, which is believed to result from X-ray irradiation
of the surface of the disk.  In our solutions, the hot boundary layer
is much thicker than the inner disk, so the disk surface should
intercept a reasonable fraction of the X-ray emission.  This fraction
will depend on the boundary layer thickness and radial extent and the
disk height profile.  Fig. 1 shows the height profile of the boundary
layer and disk for four solutions.  The solution with $\dot M =
10^{-8} \msyr$ has a thicker boundary layer and disk than the $\dot M
= 10^{-9} \msyr$ or $10^{-10} \msyr$ solutions, so a larger fraction
of the boundary layer emission should be intercepted by the disk.  The
solution for a rotating neutron star has a thinner boundary layer than
for the nonrotating star.  Note that the axes are chosen to emphasize
the differences between the three solutions and give a distorted
impression of the shape of the disk; in fact the disk is quite thin,
with $H/R \sim 0.01-0.02$ at all radii.  Thus the disk is basically
flat, and the fraction of the boundary layer emission intercepted by
the disk will be around 25\% (Lapidus \& Sunyaev 1985).  The stellar
surface will also intercept a substantial fraction of the X-ray
emission (Popham 1997), but most of this will be reradiated back into
the hot boundary layer gas.  The X-ray flux incident on the disk and
star will result in a number of interesting effects, including
polarization (Lapidus \& Sunyaev 1985), line emission, and a ``Compton
reflection'' spectrum.

\subsection{Implications for Multicomponent Spectral Fits}

With a self-consistent picture of the dynamics and energetics of the
boundary layer region, we are now in a position to assess the
multi-component models commonly used to fit LMXB spectra.  In
particular, we can discuss the emitting regions which are present.
The first component, present in most of our solutions, is a hot,
low-density boundary layer region which cools by inverse-Compton
scattering of photons emitted from the cooler, optically thick zone
near the stellar surface.  The hot region also emits bremsstrahlung
radiation, but in our solutions this is an insignificantly small
fraction of the total emission.  The temperature and density of the
gas vary across the boundary layer, so the use of a single-temperature
Comptonizing cloud will only approximate the true emitted spectrum.
This temperature variation is quite pronounced at low $\dot M$ for
$\alpha=0.1$; for other choices of parameters the single-temperature
approximation may not be so bad.

The second component is the disk, which is optically thick due to the
combined effects of absorption and scattering.  The scattering opacity
is much larger than the absorptive opacity in our solutions, so the
disk should emit a modified blackbody spectrum.  In general the
effective temperature of the inner disk is in the range $3-6 \times
10^6$ K, so $kT \simeq 0.25-0.5$ keV, but the color temperature of the
disk radiation will be higher due to the modified blackbody spectrum.
In multi-component models, the ``multi-color disk'' component is often
parameterized using an inner radius $r_{in}$; in this context it is
useful to note that the inner disk radius in our solutions is set by
the radial extent of the boundary layer, and it varies by a factor of
two as $\dot M$ changes.

Two-component models which consist of a Comptonized blackbody spectrum
plus a modified blackbody multi-color disk would provide the best
approximation to the solutions shown here.  The two components would
be constrained to having the appropriate luminosities.  For a narrow
boundary layer around a non-rotating star, the disk and boundary layer
each contribute half of the total accretion luminosity $L_{acc} \equiv
G M \dot M / R_*$.  The disk luminosity can be substantially lower
than this, since the inner radius of the disk can be as much as $\sim
2 R_*$.  The boundary layer luminosity is much less than $L_{acc}/2$
if the accreting star is rotating in the same direction as the disk,
which it should be due to accretion spinup.  The total luminosity of
the boundary layer and disk varies as $1 - jf + 0.5f^2$, where $f
\equiv \Omega_* / \Omega_K(R_*)$ is the spin rate of the star as a
fraction of the breakup rate (PN95).  If the outer edge of the
boundary layer is at a radius $b R_*$, then $j \simeq b^{1/2}$.  The
disk and boundary layer luminosities will then be
\begin{equation}
L_{disk} \simeq L_{acc} \left( {1 \over 2b} \right) \qquad L_{bl}
\simeq L_{acc} \left(1 - b^{1/2} f + {f^2 \over 2} - {1 \over 2b}
\right).
\end{equation} 
By using two components with the correct temperatures and optical
depths and these luminosities, it should be possible to produce a
reasonable approximation to the spectrum that would be emitted by our
solutions.

\subsection{Comparison with Boundary Layers in CVs and Implications for
Oscillations}

In many respects the boundary layer solutions for accreting neutron
stars in this paper resemble the solutions for very hot boundary
layers in CVs presented by NP93.  In both CVs and LMXBs, the radial
extent of the boundary layer can be comparable to the stellar radius.
One difference between the two types of solutions is that the radial
extent of the boundary layer generally increases with increasing $\dot
M$ for LMXB solutions, while it increases with {\it decreasing} $\dot
M$ for CV solutions.  Both types of boundary layers reach high
temperatures $T \sim 10^8$, but in some cases the LMXB solutions get
quite a bit hotter than this.  One might expect the LMXB boundary
layers to be much hotter than the CV ones, but as discussed above,
Compton cooling limits the temperature.  The CV boundary layer
solutions of NP93 have $\tau_s < 1$ and $T_e \sim 10^8$ K, so the
effects of Compton cooling should be small.

One important observational characteristic of LMXBs is the presence of
kHz QPOs.  One of us has argued that kHz QPOs in LMXBs are very
similar to dwarf nova oscillations (DNOs) observed in CVs, and that
DNOs could arise at the boundary between the disk and the hot,
low-density boundary layer (Popham 1999, 2000).  Since we have shown
here that LMXBs should have hot, low-density boundary layers similar
to those in CVs, the logical next step is that kHz QPOs could arise at
this same location.  In our solutions the disk--boundary layer
transition occurs at $\sim 1.1 - 2.0 R_*$, moving outward as $\dot M$
increases from $10^{-10}$ to $10^{-8} \msyr$.  The Keplerian rotation
frequencies for this range of radii are $\sim 770-1880$ Hz for our
choice of the neutron star mass ($1.4 \msun$) and radius (10 km).  For
a larger neutron star radius, e.g. 13 km, the range would be $\sim
520-1270$ Hz, which matches the observed range of frequencies quite
well.

One difficulty with this picture is that unlike the CV solutions
(NP93), our LMXB solutions have this transition radius increasing with
increasing $\dot M$.  If the oscillation period is just the Keplerian
period at the transition radius, it should also increase with
increasing $\dot M$.  However, in the oscillations observed thus far,
as the oscillation period increases, $\dot M$ is inferred to decrease.
Thus, if kHz QPOs are formed at the disk--boundary layer transition,
either $\dot M$ must change in the opposite sense to that inferred
from the observations, or the change in the transition radius with
$\dot M$ must be opposite to what our models predict.  Note that the
transition radius reaches a minimum for $\dot M < 10^{-9.8} \msyr$
(for $\alpha=0.1$), and then begins to move back out again as $\dot M$
decreases.  Thus for low values of $\dot M$, the sense of the
variation of the Keplerian period with $\dot M$ would agree with the
kHz QPO observations; however, these values of $\dot M$ correspond to
lower luminosities than are observed from the systems which show kHz
QPOs.  For higher values of $\alpha$, the transition radius might turn
around at a higher $\dot M$, and when we add additional physical
effects to our model, the dependence of the transition radius on $\dot
M$ may change.

\subsection{Limitations of the Model and Future Improvements}

One obvious limitation of the model used here is the use of
one-dimensional, vertically-averaged equations to model the boundary
layer region.  This region is inherently two-dimensional, and by using
one-dimensional equations, we are unable to simulate a number of
aspects of the flow of matter and radiation.  Most of the important
physical quantities are assumed to be constant with height $z$ above
the midplane, but in reality they vary with both $R$ and $z$.  The
viscous dissipation rate varies with $z$, and the processes which
produce radial energy transport will also produce vertical energy
transport.  As discussed in \S 2, the abrupt drop in the disk height
$H$ just before the gas reaches the stellar surface may also be a
consequence of our use of simplified one-dimensional equations.
Finally, the main differences between the model presented here and the
``spreading layer'' model of the flow of the gas over the neutron star
surface are different assumptions about the geometry of the gas as it
reaches the star.  By constructing a two-dimensional model, we could
eliminate a number of these problems.

Another important improvement to be made to the model is the inclusion
of general relativistic effects.  We have found infall velocities
$v_R$ which are at most $\sim 0.01 c$, but the accreting gas may reach
much larger infall velocities if it falls inside the marginally stable
orbit.  In addition to purely dynamical effects, radiation drag can
remove angular momentum from the gas (Miller \& Lamb 1993, 1996).

Our current equations for radiative transfer are also rather crude.
In particular, the assumption of frequency independence makes our
treatment of absorption and of Compton scattering very approximate.
Also, in our current form of the radiative transfer equations, all
radial flux stays inside the disk, except that which is scattered or
absorbed and reemitted as vertical flux.  This is reasonable when the
disk height varies slowly with radius, but in our boundary layers the
disk height varies rapidly (see Fig. 1).  This should be taken into
account; for instance, some of the outward radial flux should escape
from the outer side of the boundary layer as ``vertical'' flux when
the disk height drops rapidly there.  Since radiation pressure is
important in supporting the gas, this could affect the size of the
boundary layer region.

\newpage

\figcaption{The vertical pressure scale height $H$ (top) and the
angular velocity $\Omega$ (bottom), for solutions with $\dot M =
10^{-10}$ (long-dashed), $10^{-9}$ (solid) and $10^{-8} \msyr$
(dotted), all for a non-rotating neutron star, and for $\dot M =
10^{-9} \msyr$ and a neutron star rotation frequency $f_* = 636$ Hz
(dashed), all with standard viscosity and $\alpha=0.1$.  Note the very
small values of $H$ at the ``neck'' between the disk and boundary
layer in the lower $\dot M$ solutions, and rapid increase in $H$ in
the boundary layer.}

\figcaption{The Reynolds number $Re = \Omega R (R-R_*) / \nu_{rad}$
calculated from our solutions, but assuming that the viscosity is
given by the radiative viscosity $\nu_{rad}$.}

\figcaption{Angular velocity $\Omega$ (solid) and Keplerian angular
velocity $\Omega_K$ (dashed), radial velocity $v_r$ (solid) and sound
speed $c_s$ (dashed), electron temperature $T_e$, effective
temperature $T_{eff}$, density $\rho$, vertical pressure scale height
$H$, vertical scattering optical depth $\tau_s$, and vertical
effective optical depth $\tau_*$, for a solution with $\dot M =
10^{-9} \msyr$, $\alpha=0.1$, $\Omega_* = 0$, and standard viscosity.}

\figcaption{Additional quantities for the solution shown in Fig. 3.
(a) Total pressure (solid) and radiation pressure (dashed)in cgs
units. (b) The fraction $\beta$ of the total pressure which is due to
gas pressure. (c) Mean intensity of radiation $u$ (solid) and
blackbody intensity $B = (\sigma / \pi) T_e^4$ (dashed) in cgs
units. (d) Electron temperature $T_e$ (solid), photon temperature
$T_{phot}$ (dashed), and ion temperature $T_i$ (dotted).  (e) The
radial flux of radiation at the disk midplane (solid) and the vertical
flux at the disk surface (dotted), in units of $10^{25} \ecps$. (f)
The ratio of the radial to the vertical flux, expressed as the angle
$\tan^{-1} (F_R/F_V)$ in degrees (solid), so that zero corresponds to
vertical and 90 degrees to radial.  The small arrows show the angle
graphically, and the dotted line is the disk height $H$ (units shown
on right axis).}

\figcaption{(Top row) Radial acceleration due to gravity (solid),
centrifugal acceleration (dotted), pressure gradient (dashed), bulk
viscosity (long-dashed), and radial velocity gradient (dot-dashed) in
units of $10^{14} \cm~ \s^{-2}$.  (Second row) Energy gain or loss per
unit disk midplane area in the outer boundary layer: advection
(solid), Comptonization (long-dashed), viscous dissipation (dotted),
radiation (short-dashed), in units of $10^{25} \ecps$.  (Third row)
Viscous transport of energy in the boundary layer and disk (upper
panels), where positive values correspond to outward radial transport.
(Bottom row) The energy deposited or removed from the disk, normalized
by the local gravitational energy release (lower panels), by viscous
dissipation (solid), change in rotational kinetic energy (dotted), and
viscous transport (dashed).}

\figcaption{The energy deposited or removed from the disk, normalized
by the local gravitational energy release, as a function of (top) the
radial optical depth measured along the midplane, and (bottom) the
vertical optical depth from the midplane to the surface.  For the
radial optical depth, $\tau = 0$ corresponds to the point where
$d\Omega / dR = 0$, and $\tau$ is increasing inward.  The upper axes
shows some corresponding values of the radius $R$.  Due to the viscous
transport of energy, energy is dissipated where $\tau$ is large.}

\figcaption{The change in the boundary layer structure as a function
of the mass accretion rate $\dot M$ for five solutions with $\dot M =
10^{-8}$, $10^{-8.5}$, $10^{-9}$, $10^{-9.5}$, and $10^{-10} \msyr$.}

\figcaption{The Eddington radial flux (dotted) and the radial flux for
the $\dot M = 10^{-8} \msyr$ (solid), $10^{-9} \msyr$ (short-dashed),
and $10^{-10} \msyr$ (long-dashed) solutions.}

\figcaption{The change in the boundary layer structure with $\dot M$,
for values of $\dot M$ near the Eddington limit: $dot M = 10^{-8}$,
$10^{-7.95}$, $10^{-7.90}$, and $10^{-7.85} \msyr$.}

\figcaption{Same as Fig. 7, but for $\alpha=0.01$.}

\figcaption{(a) The turbulent length scale $l_{turb} = (H^{-2} +
H_r^{-2})^{-1/2}$ (solid), the vertical pressure scale height $H$
(dotted), the radial pressure scale height $H_r$ (short-dashed), and
the subsonic length scale $H_s$ (long-dashed) for the solution with
$\dot M = 10^{-9} \msyr$, $\alpha=0.1$, $\Omega_* = 0$, and standard
viscosity.  In the boundary layer, $l_{turb} > H_s$. (b) Same as (a),
but for a solution with subsonic viscosity.}

\figcaption{Same as Fig. 7, but for ``subsonic'' viscosity (see text)
with $\alpha = 0.1$.}

\figcaption{Comparison of the electron temperature $T_e$ and the
vertical scattering optical depth $\tau_s$ for solutions with regular
and subsonic viscosity.  Solutions in the top panels have $\dot M =
10^{-9} \msyr$ and subsonic viscosity with $\alpha = 0.1$ (dotted),
0.2 (dashed), or standard viscosity with $\alpha=0.1$ (solid); those in
the bottom panels have $\dot M = 10^{-10} \msyr$ and subsonic
viscosity with $\alpha = 0.1$ (dotted), 0.316 (dashed), or regular
viscosity with $\alpha=0.1$ (solid).}

\figcaption{The change in the boundary layer structure with the
rotation rate of the accreting neutron star for five solutions with
$f_* = 318$, 637, 1273, 1910 Hz ($\Omega_* = 0$, 2000, 4000, 8000,
and 12000 $\pers$. $\dot M = 10^{-9} \msyr$ and $\alpha = 0.1$ for all
solutions.  The dashed line is a solution with $f_* = 2085$
Hz $(\Omega_* = 13100 \pers)$ and $j=0$.}

\begin{figure}
\plotone{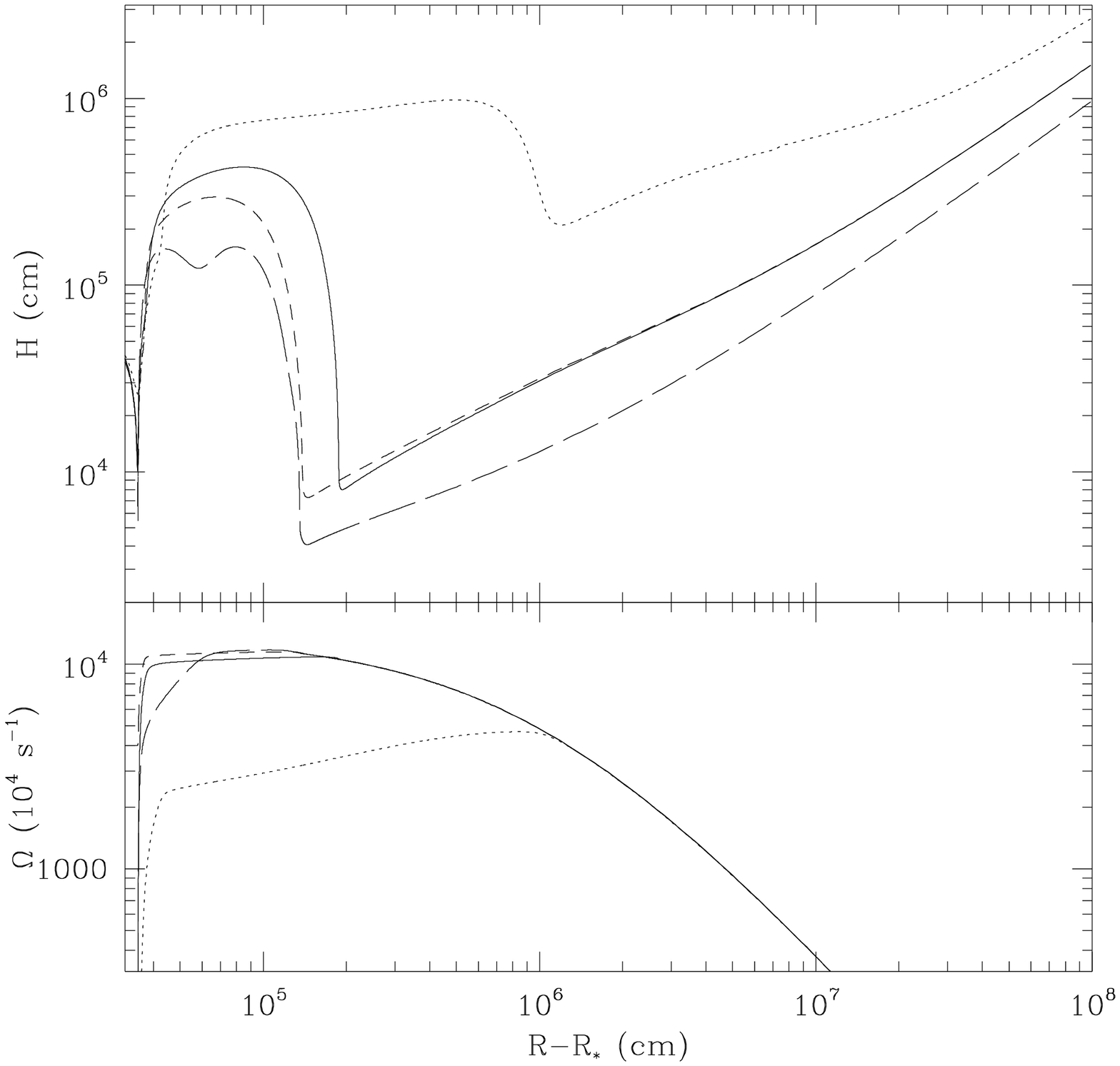}
\end{figure}
\begin{figure}
\plotone{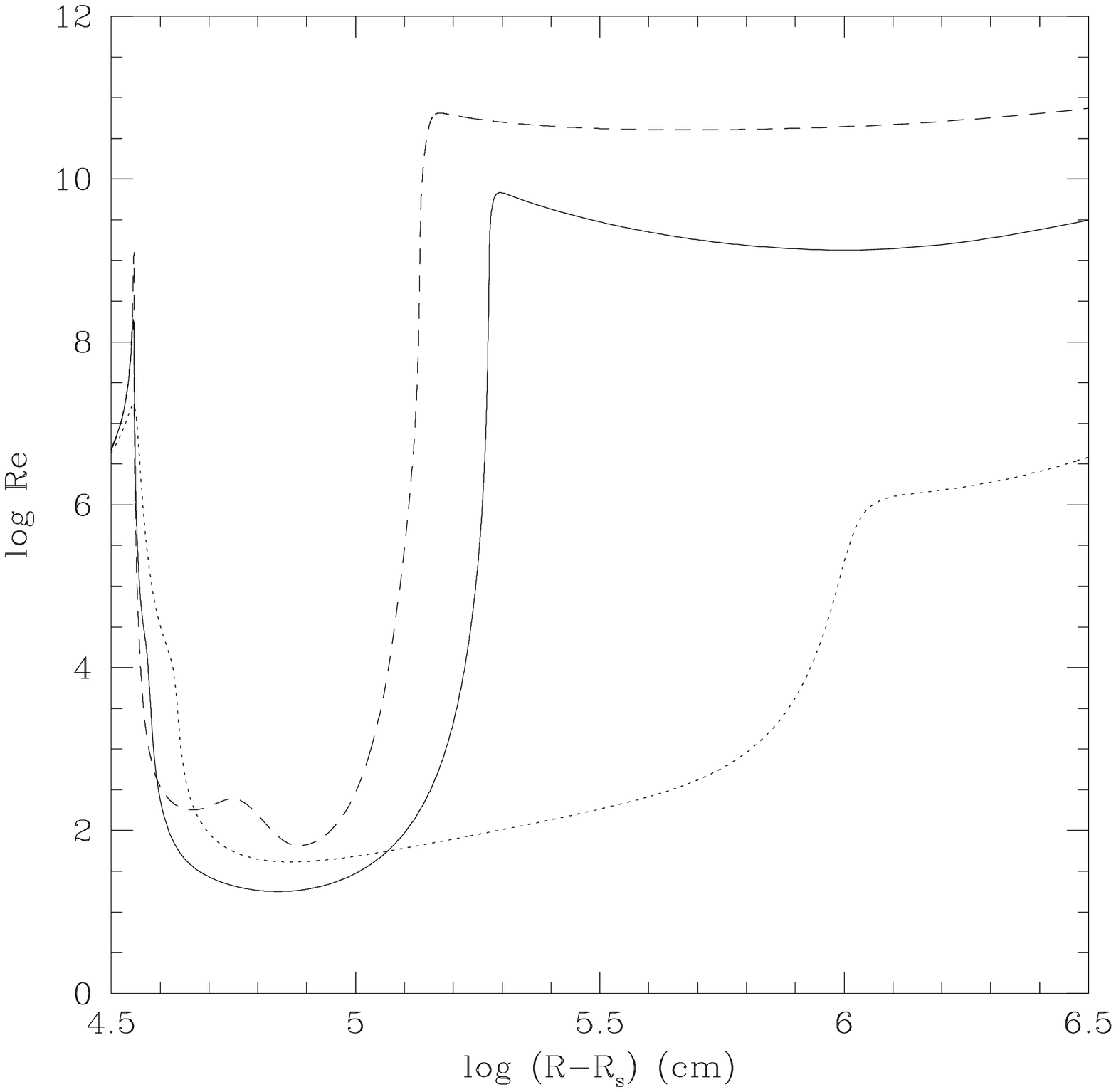}
\end{figure}
\begin{figure}
\plotone{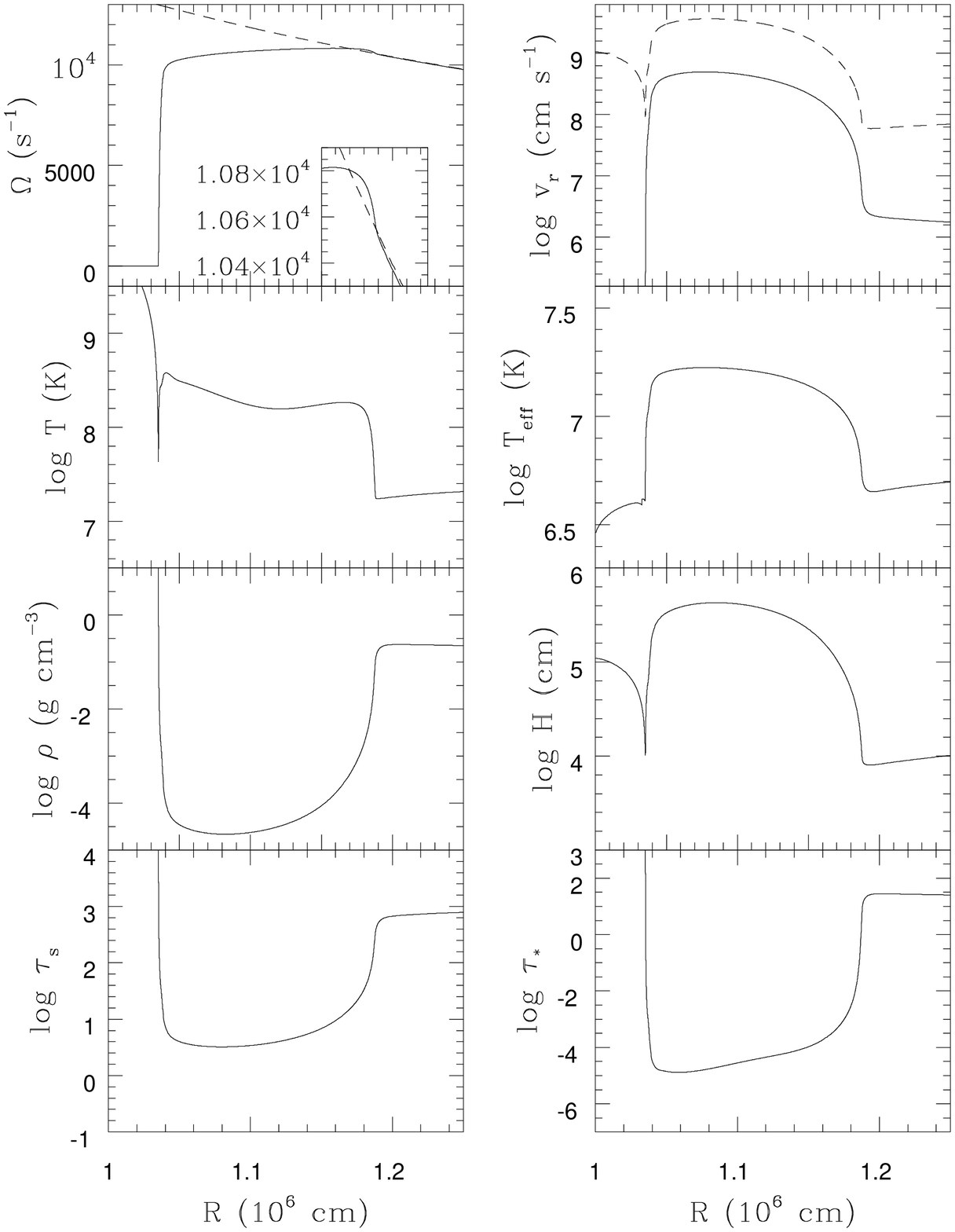}
\end{figure}
\begin{figure}
\plotone{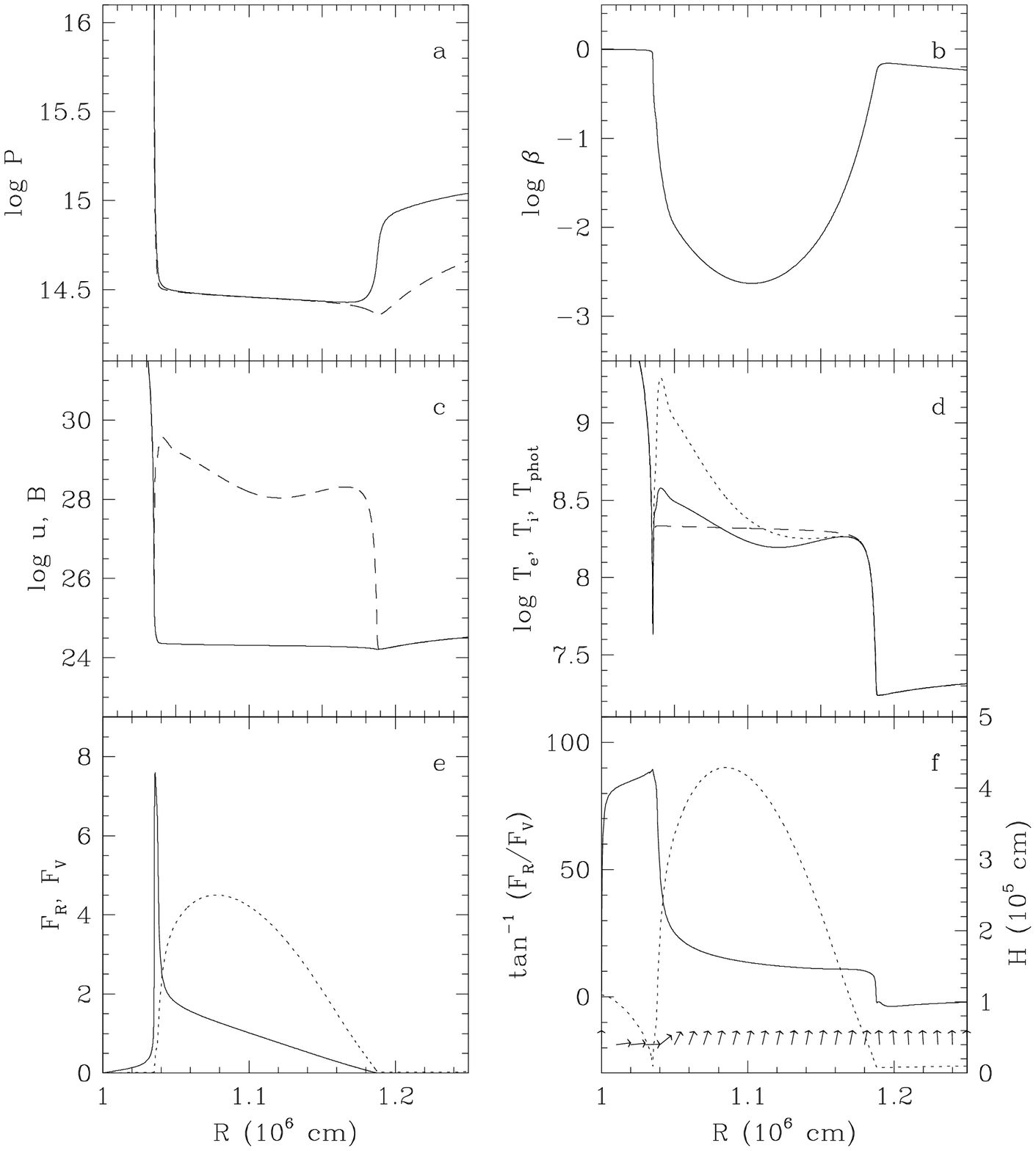}
\end{figure}
\begin{figure}
\plotone{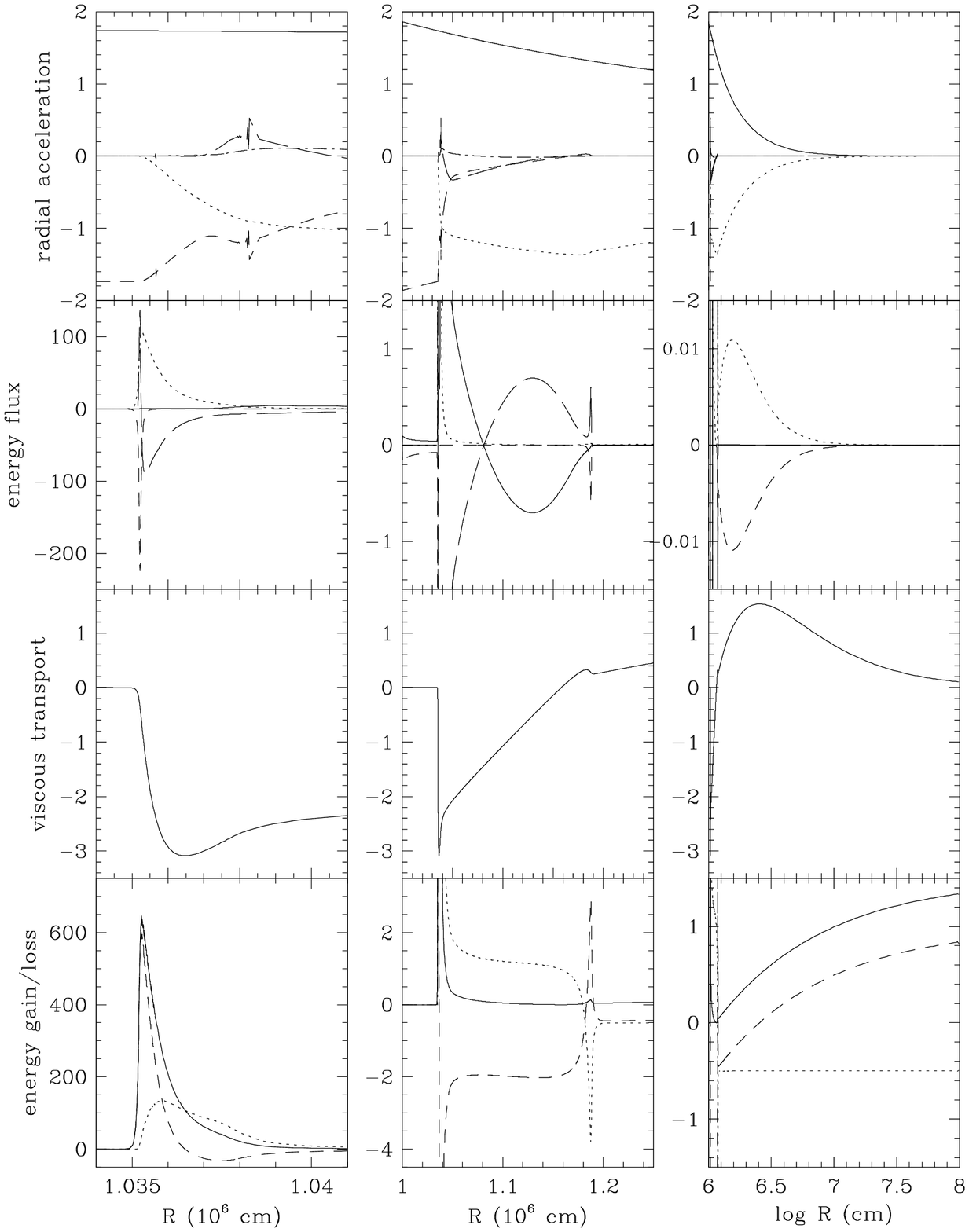}
\end{figure}
\begin{figure}
\plotone{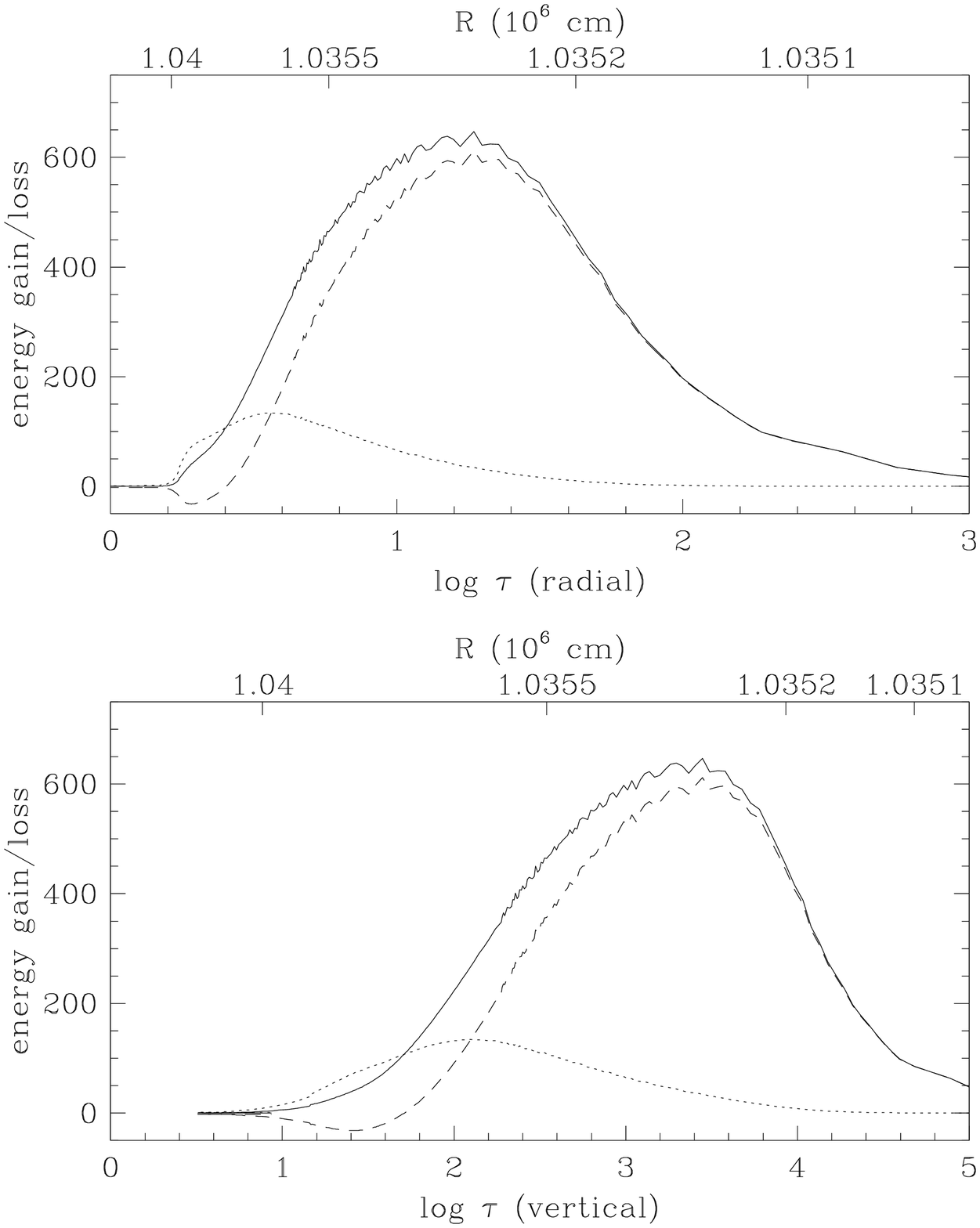}
\end{figure}
\begin{figure}
\plotone{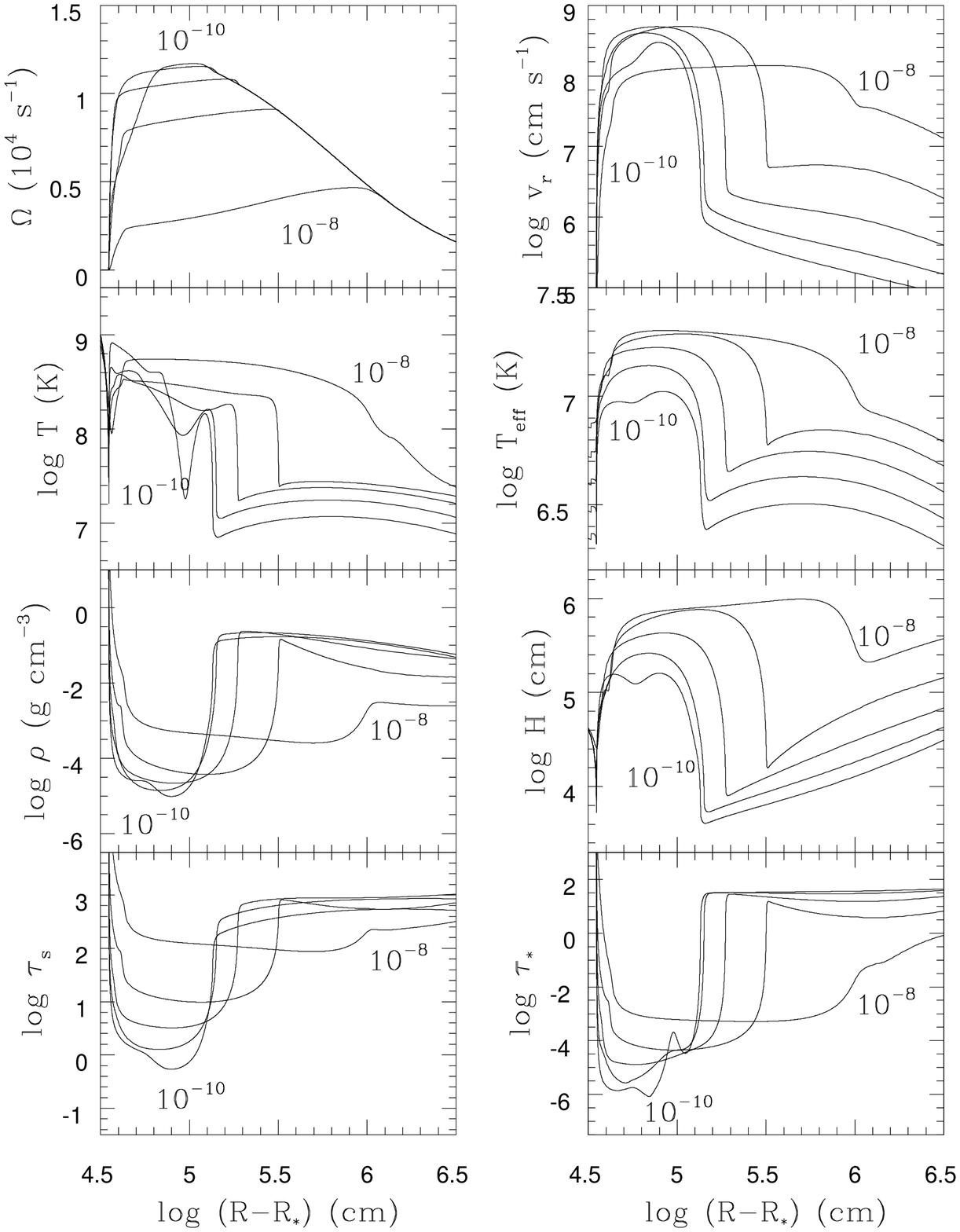}
\end{figure}
\begin{figure}
\plotone{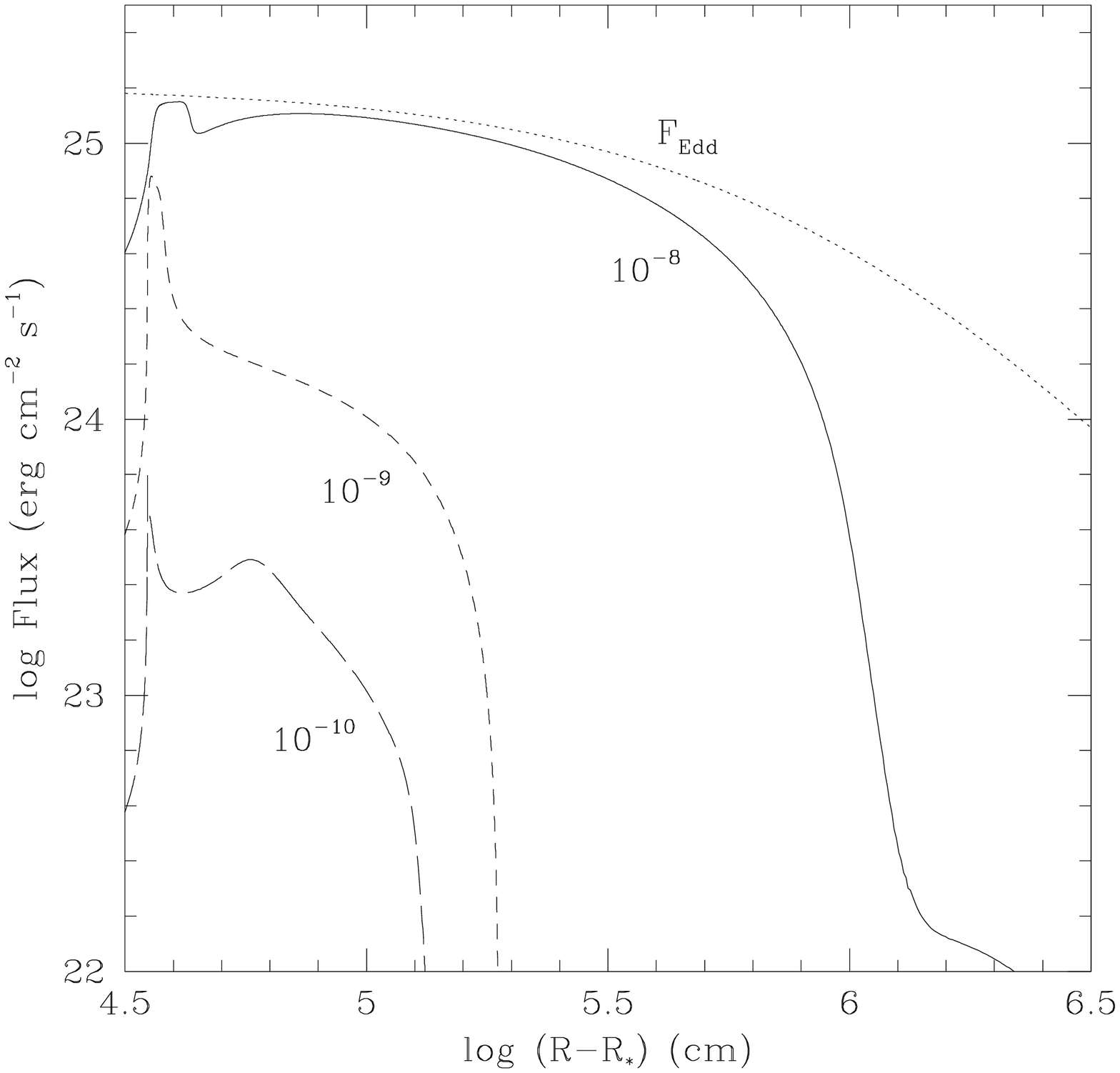}
\end{figure}
\begin{figure}
\plotone{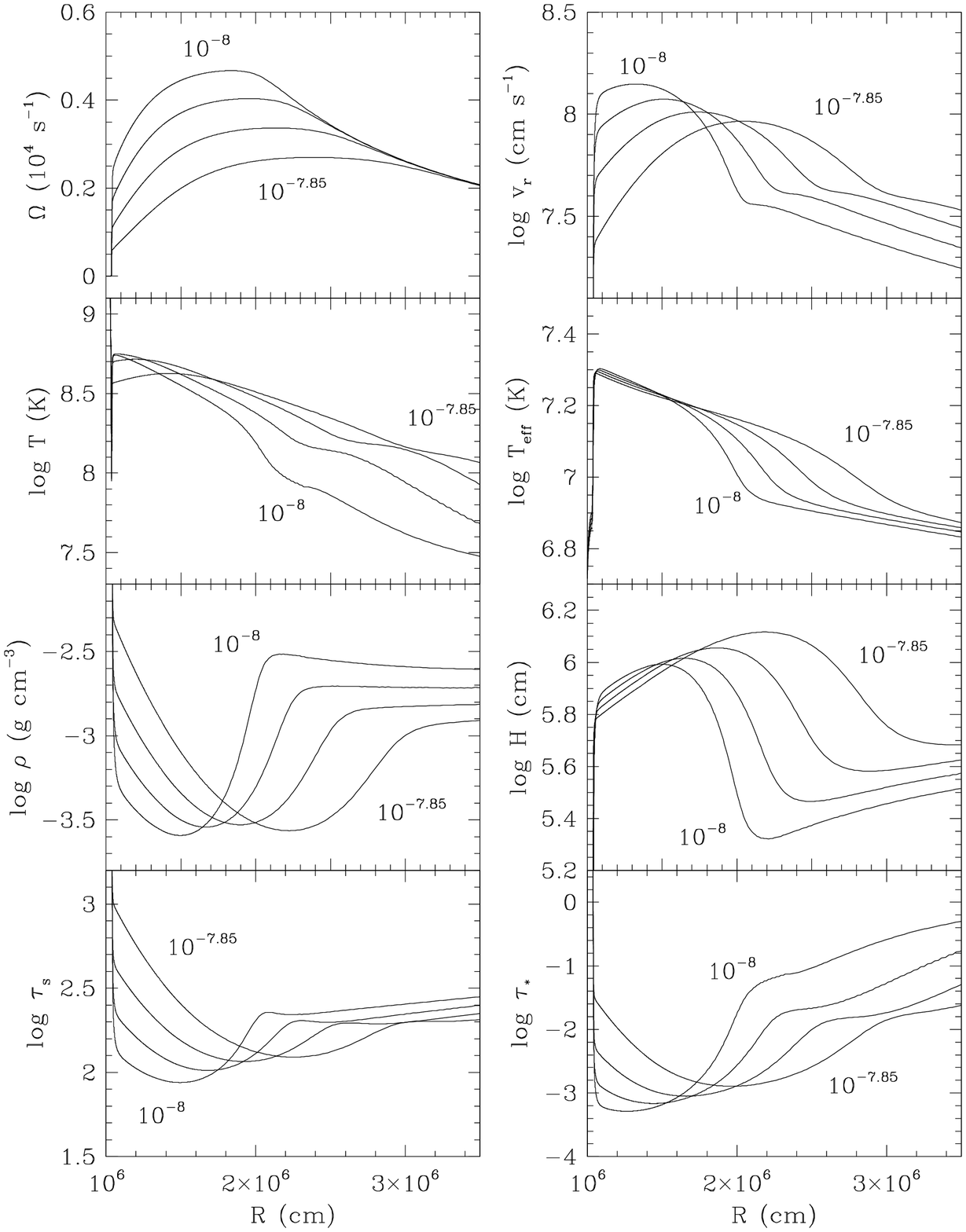}
\end{figure}
\begin{figure}
\plotone{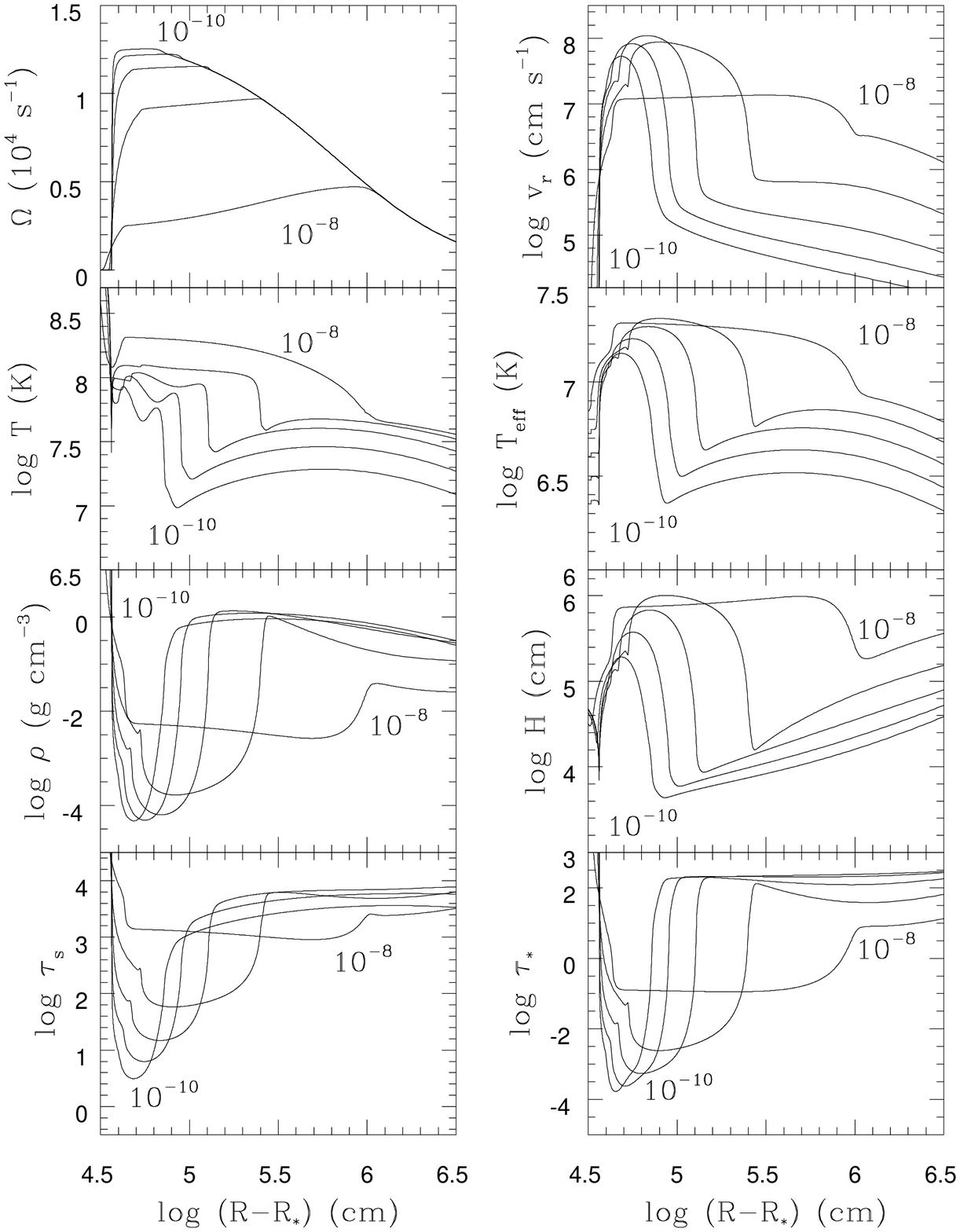}
\end{figure}
\begin{figure}
\plotone{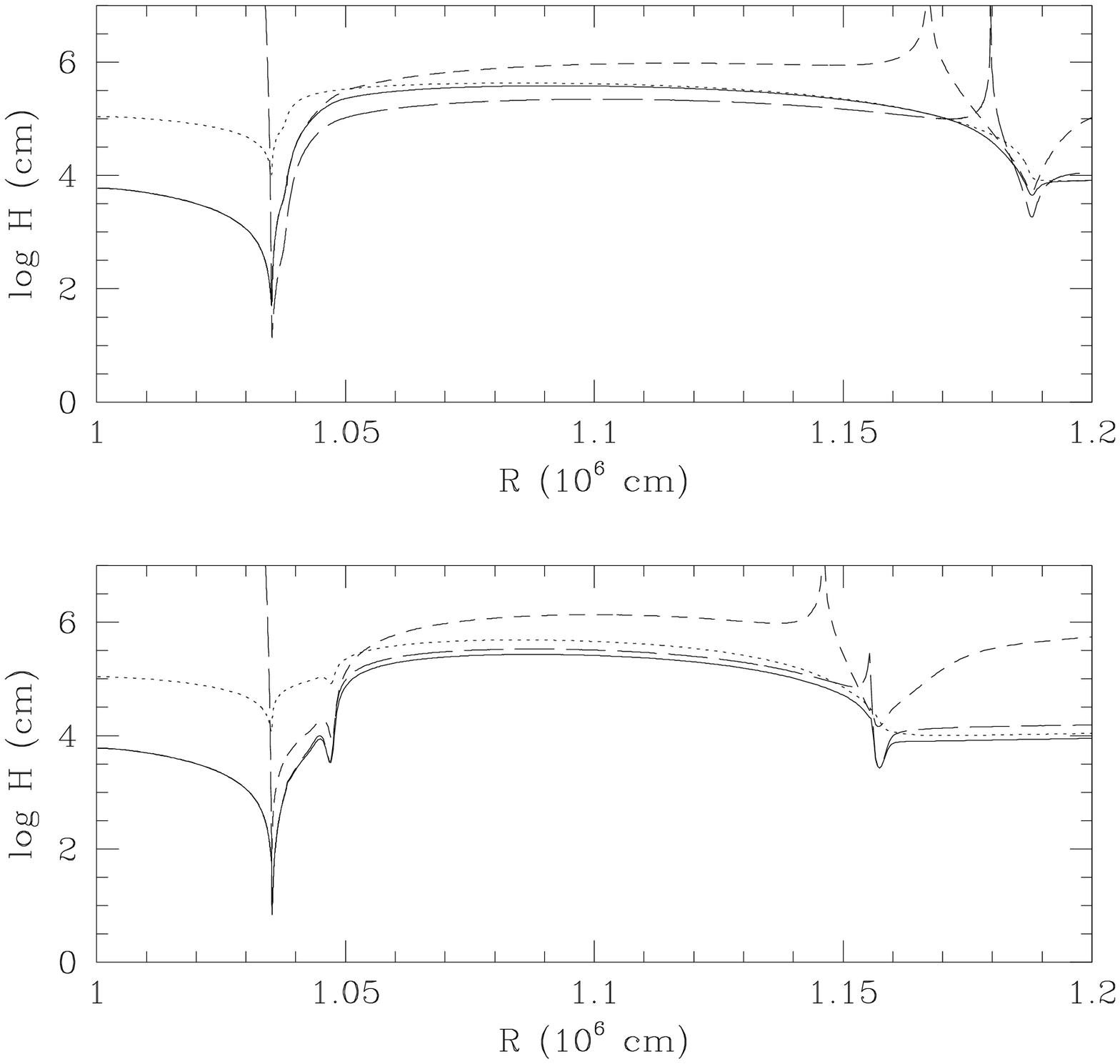}
\end{figure}
\begin{figure}
\plotone{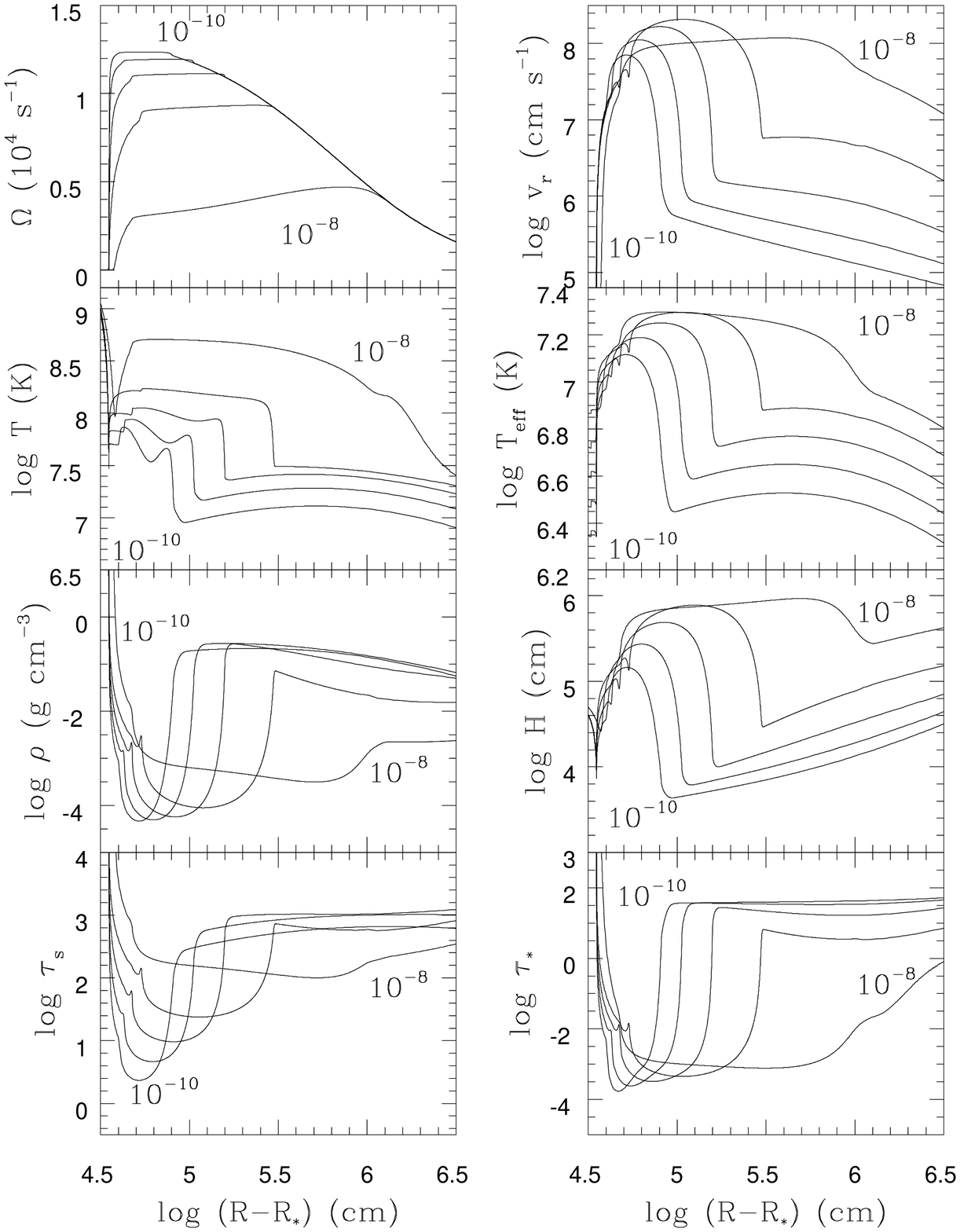}
\end{figure}
\begin{figure}
\plotone{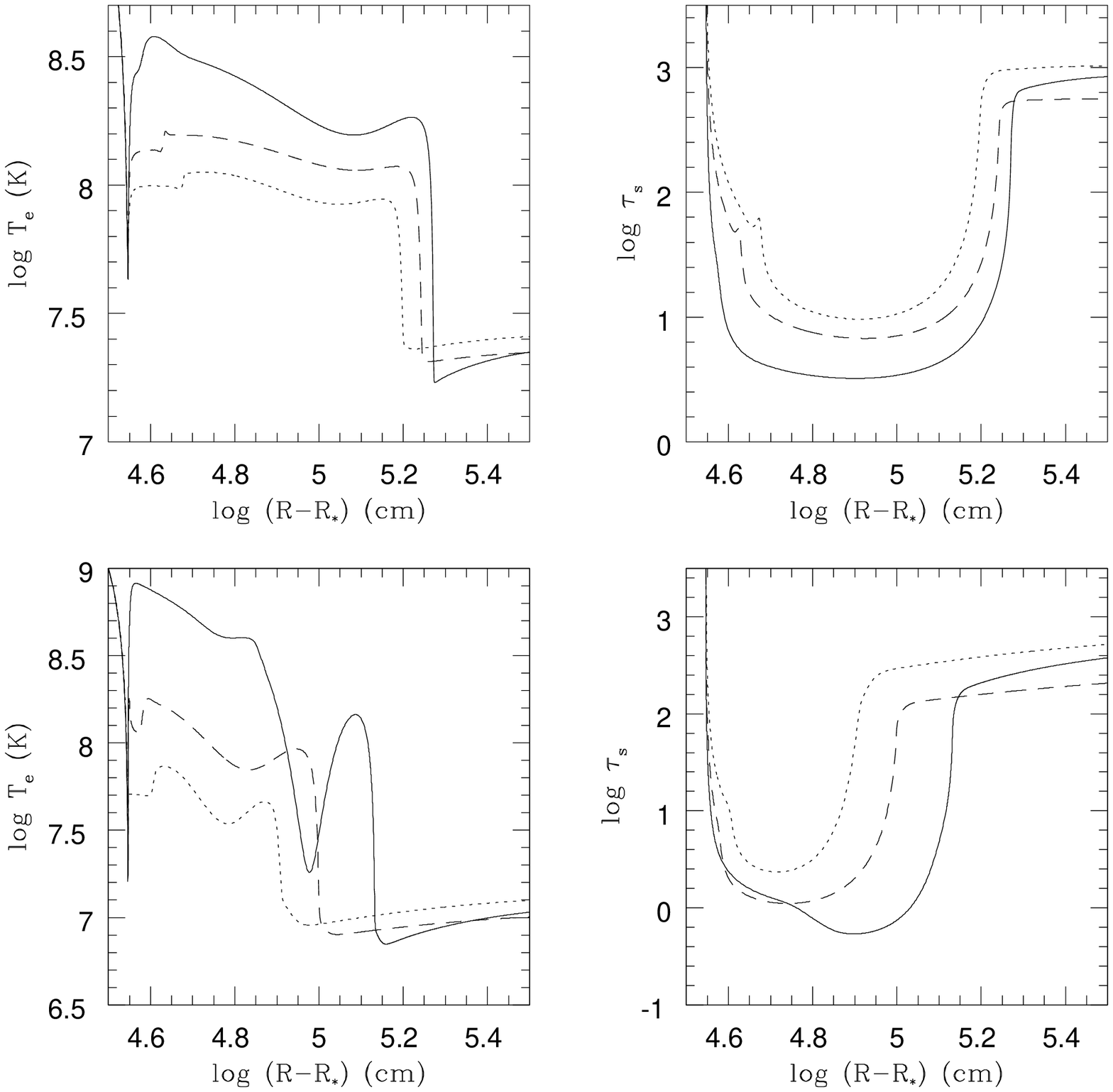}
\end{figure}
\begin{figure}
\plotone{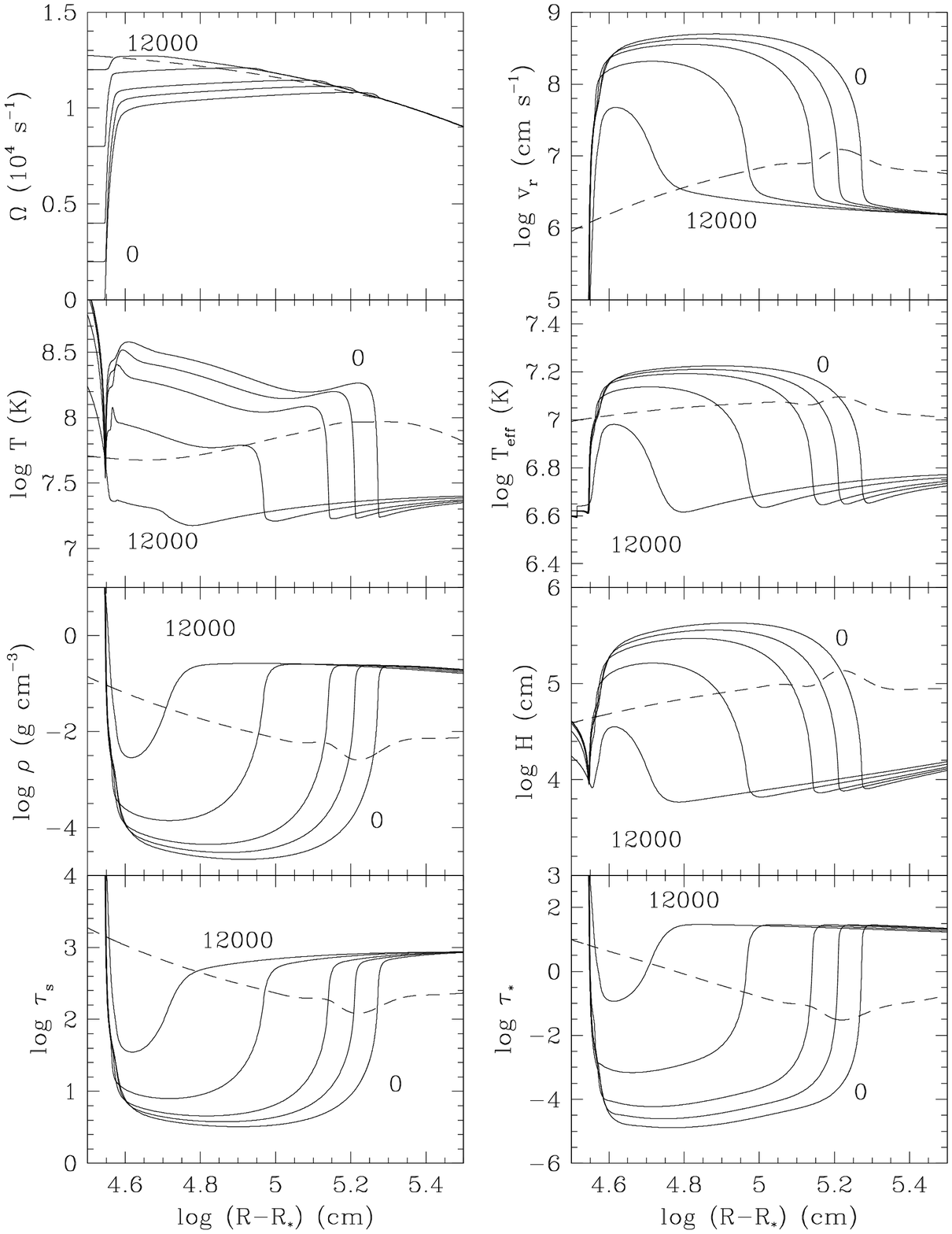}
\end{figure}

\end{document}